\documentclass[10pt]{article}
	\usepackage{caption}
	\usepackage{url}
	\usepackage{verbatim}
    \usepackage{graphicx}
    \usepackage{tikz}
    \usepackage{amsmath}
	\usepackage{mathrsfs}
	\usepackage{mathtools}
	\usepackage{stmaryrd}
    \usepackage{amssymb}
    \usepackage{amsthm}
    \usepackage{url}
    \usepackage{setspace}
    \usepackage{longtable}
    \usepackage{cmbright}
    \usepackage{color}
    \usepackage{fancyhdr}
    \usepackage{nicefrac}
	\usepackage{enumerate}

    \newcommand{\bfa}{\boldsymbol{a}}

    \newcommand{\bfJ}{\boldsymbol{J}}

    \newcommand{\bfn}{\boldsymbol{n}}

    \newcommand{\bfx}{\boldsymbol{x}}

    \numberwithin{equation}{section}

\begin{document}

\title{A Thermodynamically-Consistent Phase Field Crystal Model of Solidification with Heat Flux}

\author{Cheng Wang\thanks{Department of Mathematics, The University of Massachusetts, North Dartmouth, MA 02747 (cwang1@umassd.edu)}  \and  Steven M. Wise\thanks{Corresponding author: Department of Mathematics, The University of Tennessee, Knoxville, TN 37996 (swise1@utk.edu)} }

	\maketitle
	\numberwithin{equation}{section}

	\begin{abstract}
In this paper we describe a new model for solidification with heat flux using the phase field crystal (PFC) framework. The equations are thermodynamically consistent, in the sense that the time rate of change of the entropy density is positive in the bulk and at the boundaries of the domain of interest. The resulting model consists of two equations, a heat-like equation and a mass-conservation equation that describes how the atom density changes in time and space. The model is simple, yet it can properly capture the variation in the free energy landscape as the temperature is changed. We describe to construct a temperature-atom-density phase diagram using this energy landscape, and we give a simple demonstration of solidification using the model.
	\end{abstract}



	\section{Introduction}
	
The phase field crystal (PFC) model was introduced in~\cite{elder02,elder04} as continuum description of solidification in a unary material. It was formulated as a mass conservative version of the classical Swift-Hohenberg equation, but, later, the model was re-derived,  via certain reasonable simplifications, from the dynamical density functional theory (DDFT)~\cite{elder07}. In particular, assuming a constant, uniform temperature field $T$, one expresses the Helmholtz free energy density via
	\[
F= \int_\Omega\left\{f(T,\rho) + \frac{T\kappa_{f,\rho,o}}{2}(\rho-\rho_o) \mathcal{C} (\rho-\rho_o) \right\}d\bfx,
	\]
where $\Omega$ is some spatial domain of interest; $\rho:\Omega \to [0,\infty)$ is the number density field of the unary material in $\Omega$; the constant $\rho_o>0$ is a reference density; $\kappa_{f,\rho,o} >0$ is a positive constant; $f$ is the homogeneous Helmholtz free energy density; and $\mathcal{C}$ is a symmetric, potentially nonlocal, two-point correlation operator. The free energy density, $f$, is often taken to satisfy an ideal ``gas" model:
	\[
f(T,\rho) = \rho k_BT \ln\left( \frac{\rho}{\rho_o}\right) -  k_BT\left(\rho -\rho_o\right)  ,
	\]
where $k_B$ is the Bolzmann constant. Often, one makes a (Taylor) polynomial approximation of the logarithmic term about the reference density to make the model more tractable. However, it is the singular nature of the logarithmic term that guarantees the positivity of the solutions, and this is an important feature in the numerical and PDE analyses.  At constant temperature, one can argue that the dynamics of the model should satisfy a diffusion-dominated mass conservation equation of the form
	\[
\dot{\rho} = -\nabla\cdot \bfJ , \quad \bfJ = -M \rho \nabla \mu,	
	\]
where $\bfJ$ is the diffusion flux; $M>0$ is a mobility; and $\mu$ is the chemical potential:
	\[
\mu := \delta_\rho F = k_BT\log (\rho) + T \kappa_{f,\rho,o} \mathcal{C} (\rho-\rho_o) ,
	\]
where we have assumed, for simplicity, that the boundary conditions are periodic. As a consequence of these assumptions, the total free energy is dissipated as the system evolves toward equilibrium, and the dissipation rate is
	\[
\dot{F} = - M\int_\Omega \rho|\nabla \mu|^2\, d\bfx \le 0.
	\]
Of course, it would be necessary to justify the property that $\rho >0$ (or at least $\rho\ge0$) point-wise for the model to make sense. Numerical analyses of similar gradient flow models, that is, models that have logarithmic energy potentials, have been performed in~\cite{chen19b,Dong2020b,  dong19b, dong20a, LiuC2021b, LiuC2020, QianWangZhou_JCP20, Yuan2021a}.

The PFC modeling framework has a couple of basic, distinctive features. First, the solutions to the PFC-type models exhibit (at least) two distinct phases. One is a spatially oscillatory phase, which is identified with the solid phase, and the other is a spatially uniform phase, which is usually identified as the liquid (or gas) phase. The peaks of the solutions in the oscillatory phase are interpreted as the ``locations" of the atoms, and typically, one can choose $\mathcal{C}$ so that the peaks are arranged in a desired crystal structure~\cite{provatas07}. Second, PFC models operate at atomic length scales but diffusive time scales. Thus, the framework can capture long-time phenomena.
	
In this paper, we will devise a new model for solidification and melting using the phase field crystal framework. In particular, we will not assume that the temperature is uniform in space and time. For an adiabatically isolated system, this requires that the global entropy is increasing as the system moves towards equilibrium, while, at the same time, the total energy and the number density of particles should be conserved.  From a modeling perspective, one has some freedom in choosing the equations so that these properties hold. In contrast with the case of uniform temperature, the Helmholtz free energy can increase or decrease in time. While important to the model formulation, it alone does not dictate the drive of the system toward equilibrium.  

While our model is, to our knowledge, new and distinct, an earlier effort has made some important contributions towards adding heat flux to the PFC framework for melting/solidification. Specifically, in the paper by Kocher and Provatas~\cite{Kocher2019}, the authors add temperature variation in the study of rapid solidification. In fact, the framework that they derive is quite general, and, as they claim, seems to apply to several physical settings.  However, the equations that they derive are ultimately different from those found herein. Our model is  simpler in the description of the internal energy and the latent heat, but it still captures the most important effects of temperature variation. In particular, the heat equation that we derive is essentially linear. Ultimately, Kocher and Provatas introduced a couple of physically reasonable approximations in the derivation of their working model. In particular, they have introduced a smoothing operation to the internal energy density. The result of these approximations is that their working model is no longer provably entropy non-increasing.  Our model, on the other hand, retains this important property.

In this brief paper, we will construct and demonstrate the key features of this thermal-PFC model. We plan more thorough numerical, mathematical, and physical investigations in future works.  The paper is arranged as follows. In Section~\ref{sec:derivation}, we derive the model using the framework outlined in classical phase field community in the 1990s. In Section~\ref{sec:phase-diagrams}, we show how to construct phase diagrams from the model. Finally, we use the free energy landscape and the phase diagram information to do some very simple computations in Section~\ref{sec:numerics}.

	\section{Derivation}
	\label{sec:derivation}
	
The derivation here follows the ideas in the now classic papers by Charach and Fife~\cite{charach1998}, Wang~\emph{et al.}~\cite{wang1993}, and Wheeler, McFadden, and Boettinger~\cite{wheeler1996}. These papers described thermodynamically consistent phase field models of solidification with heat flux, that is, with a non-uniform and variable temperature field. The main difference between these classical solidification models and and PFC modeling framework is that, in the latter, it is a non-trivial task to identify the equilibrium phases. Indeed, in the PFC framework, great care must be taken in identifying the equilibrium liquid and solid states via the free energy landscape.
	
	\subsection{Basic Assumptions}
	
Set $\Omega\subset\mathbb{R}^d$ and let $e, s, f:\Omega\to\mathbb{R}$ denote the internal energy, entropy, and Helmholtz free energy densities (per unit volume). The functions $T, \rho:\Omega\to \mathbb{R}$ are the temperature and the number density (of particles) fields of a unary material occupying the volume $\Omega$.  The densities $e$, $s$, and $f$ are functions only of the local values of the thermodynamic variables and do not depend upon gradients, and, for this reason, they are usually called homogenous energy and entropy densities.

We assume that the total free energy, entropy, and internal energies have the following forms, respectively:
	\[
F= \int_\Omega\left\{f(T,\rho) + \frac{\kappa_{f,\rho}}{2}(\rho-\rho_o) \mathcal{C} (\rho-\rho_o) \right\}d\bfx 	,
	\]
	\[
S = \int_\Omega\left\{s(e,\rho)  - \frac{\kappa_{s,\rho}}{2}(\rho-\rho_o) \mathcal{C} (\rho-\rho_o) \right\}d\bfx 	,
	\]
	\[
E = \int_\Omega\left\{e(s,\rho)  + \frac{\kappa_{e,\rho}}{2}(\rho-\rho_o) \mathcal{C} (\rho-\rho_o) \right\}d\bfx ,
	\]
where $\mathcal{C}$ is a long-range interaction, or correlation operator, and $\rho_o>0$ is a reference density. In this paper, we will assume that $\mathcal{C}$ is a differential operator of the form
	\[
\mathcal{C}\varrho = L^{-2} \gamma \varrho + 2\Delta \varrho +L^2 \Delta^2 \varrho,
	\]
where $\gamma$ is a dimensionless parameter, and $L$ is a characteristic length. In fact, we could instead assume that $\mathcal{C}$ is a nonlocal operator, as has been done in the classical density functional theory~\cite{elder07,Kocher2019}. We require that $F = E-TS$ (globally) and $f = e - T s$ (locally) , which implies that
	\[
\kappa_{f,\rho} = \kappa_{e,\rho} +T \kappa_{s,\rho} .
	\]
Our model could be greatly simplified, by choosing  $\kappa_{e,\rho} = 0$, as was done in~\cite{wang1993,wheeler1996}, which yields
	\[
\kappa_{f,\rho} = T \kappa_{s,\rho} .
	\]
It is assumed that $\kappa_{f,\rho}$ is linear in temperature: $\kappa_{f,\rho} = \kappa_{f,\rho,o} T$, where $\kappa_{f,\rho,o} >0$ is a constant. Thus
	\begin{align*}
F & = \int_\Omega\left\{f(T, \rho)  + \frac{\kappa_{f,\rho,o} T}{2}(\rho-\rho_o) \mathcal{C} (\rho-\rho_o) \right\}d\bfx 	,
	\\
S & = \int_\Omega\left\{s(e,\rho) - \frac{\kappa_{f,\rho,o}}{2}(\rho-\rho_o) \mathcal{C} (\rho-\rho_o) \right\}d\bfx 	,
	\\
E & = \int_\Omega\left\{e(s, \rho)  \right\}d\bfx 	.	
	\end{align*}
	
	\subsection{Entropy Production}

For the evolution equations, we appeal to conservation laws and entropy production requirements, assuming those processes are diffusion dominated. Because energy is conserved locally and, typically, globally, we have the equation
	\[
\dot{e} = -\nabla \cdot \bfJ_e .
	\]
It is expected that this will ultimately provide an equation for the temperature, $T$. Since the number of particles should be conserved locally and, typically, globally, we employ for $\rho$ a mass conservation equation of the form 
	\[
\dot{\rho} = -\nabla \cdot \bfJ_\rho .
	\]
Now, we want the entropy to increase locally and globally~\cite{deGrootMazur62}. To this end, we calculate the time derivative of the total entropy
	\[
\dot{S}  = \int_\Omega \left\{ \left(\frac{\partial s}{\partial e}\right)_{\rho} \dot{e} + \left(\frac{\partial s}{\partial \rho}\right)_{e} \dot{\rho} \right\} d\bfx  + \int_\Omega  \kappa_{f,\rho,o} \mathcal{C} (\rho-\rho_o) \dot{\rho}  \, d\bfx ,
	\]
upon assuming local thermodynamic equilibrium (LTE) boundary conditions for $\rho$, so that integration-by-parts could be carried out without the introduction of any boundary integrals. In essence, LTE boundary conditions ensure that $\mathcal{C}$ is a symmetric operator: $\int_\Omega \varphi \mathcal{C}\varrho\, d\bfx = \int_\Omega \varrho \mathcal{C}\varphi\, d\bfx$.

We assume that the internal energy density is, in its most natural form, a twice continuously differentiable function of $s$ and $\rho$. We further assume that the system evolves in such a way that it never deviates greatly from equilibrium, and, consequently, we can use the equilibrium thermodynamic theory to develop our equations.   The first and second laws are encoded in the Gibbs relation
	\[
de = \left(\frac{\partial e}{\partial s}\right)_{\rho} d s  + \left(\frac{\partial e}{\partial \rho}\right)_{s } d \rho .
	\]
By definition, the temperature satisfies $\left(\frac{\partial e}{\partial s}\right)_{\rho} =: T >0$. Thus,
	\[
de = T d s + \left(\frac{\partial e}{\partial \rho}\right)_{s } d \rho .
	\]
As we have seen, the Helmholtz free energy density results from a Legendre transformation of the other densities, namely, $f = e - T s$. Thus, it follows that
	\[
df = de - T ds - s dT =  -s dT  + \left(\frac{\partial e}{\partial \rho}\right)_{s} d \rho .
	\]
The natural variables of the free energy density, $f$, are $T$ and $\rho$, as we have indicated above. This implies the Maxwell relation
	\begin{equation}
\left(\frac{\partial f}{\partial \rho}\right)_{T} = \left(\frac{\partial e}{\partial \rho}\right)_{s} .
	\label{eqn:Maxwell-1}
	\end{equation}
We want to choose our diffusion fluxes so that $0 \le \dot{S}$. Observing that
	\[
ds = \frac{1}{T} de  - \frac{1}{T}\left(\frac{\partial e}{\partial \rho}\right)_{s} d \rho ,
	\]
it follows that
	\[
\left(\frac{\partial s}{\partial e}	\right)_{\rho} = \frac{1}{T}, \quad \left(\frac{\partial s}{\partial \rho}	\right)_{e} = - \frac{1}{T}\left(\frac{\partial e}{\partial \rho}\right)_{s} .
	\]
Thus, we can write
	\[
\dot{S} = \int_\Omega \left\{  -\frac{1}{T} \nabla\cdot \bfJ_e  + \frac{1}{T}\left(\frac{\partial e}{\partial \rho}\right)_{s } \nabla\cdot \bfJ_\rho \right\} d\bfx  + \int_\Omega  \kappa_{f,\rho,o} \mathcal{C} (\rho-\rho_o) \nabla\cdot \bfJ_\rho  d\bfx .
	\]
Using the Maxwell relation~\eqref{eqn:Maxwell-1}, we may write an even more convenient form for the change in entropy:
	\[
\dot{S}  = \int_\Omega \left\{  -\frac{1}{T} \nabla\cdot \bfJ_e + \frac{1}{T}\left(\frac{\partial f}{\partial \rho}\right)_{T} \nabla\cdot \bfJ_\rho \right\} d\bfx  + \int_\Omega \kappa_{f,\rho,o} \mathcal{C} (\rho-\rho_o) \nabla\cdot \bfJ_\rho  d\bfx .
	\]
Next, we define the generalized chemical potential
	\[
\mu : = \frac{1}{T}\delta_\rho F =  \frac{1}{T}\left(\frac{\partial f}{\partial \rho}\right)_{T} + \kappa_{f,\rho,o} \mathcal{C} (\rho-\rho_o) .
	\]
Consequently,
	\[
\dot{S}  = \int_\Omega \left\{ \nabla \left(\frac{1}{T}\right) \cdot  \bfJ_e   -\nabla \mu \cdot \bfJ_\rho \right\} d\bfx + \int_{\partial\Omega} \left\{-\frac{1}{T}  \bfJ_e\cdot \bfn  + \mu  \bfJ_\rho\cdot \bfn  \right\} d\bfa . 
	\]

To get local and global entropy production, we make the following constitutive choices: for the fluxes,
	\begin{align*}
\bfJ_e = M_T \nabla \left(\frac{1}{T}\right), \quad \bfJ_\rho = - M_\rho \nabla \mu .
	\end{align*}
For the boundary conditions, we take non-negative entropy production conditions:
	\[
\bfJ_e \cdot \bfn = -\frac{\beta_{T,\partial\Omega}}{T} , \quad \bfJ_\rho\cdot \bfn  = \beta_{\rho,\partial\Omega} \mu \quad \mbox{on $\partial\Omega$},
	\]
where $\beta_{T,\partial\Omega},\beta_{\rho,\partial\Omega}  \ge 0$. These  boundary conditions allow for heat and mass to flow through the outer boundary, but only in an entropy non-decreasing manner. For adiabatically insulated materials (meaning no mass or heat is exchanged between $\Omega$ and the outside world), we assume that $\beta_{T,\partial\Omega} = \beta_{\rho,\partial\Omega} = 0$. Another common, outer boundary condition results from a constraint on the temperature:
	\[
T= T_\star < T_M \quad \mbox{on $\partial\Omega$},
	\]
known as the undercooling condition, where $T_M$ is the melting temperature. For this condition, we do not have control of the global entropy production.

For the (general) entropy non-decreasing boundary conditions, the entropy production rate becomes 
	\begin{equation}
\dot{S}   = \int_\Omega \left\{ M_T\left|\nabla \left(\frac{1}{T}\right) \right|^2   +M_\rho \left|\nabla \mu\right|^2 \right\} d\bfx + \int_{\partial\Omega} \left\{  \frac{\beta_{T,\partial\Omega}}{T^2} +  \beta_{\rho,\partial\Omega}  \mu^2  \right\} d\bfa \ge 0 ,
	\label{eqn:rate-entropy-prod}
	\end{equation}
and the evolution equations are 
	\begin{align*}
\dot{e} & = - \nabla\cdot\left( M_T \nabla \left(\frac{1}{T}\right) \right) ,
	\\
\dot{\rho} & =  \nabla\cdot \left(M_\rho\nabla \mu\right) ,
	\\
\mu & = \frac{1}{T}\left(\frac{\partial f}{\partial \rho}\right)_{T} + \kappa_{f,\rho,o} \mathcal{C} (\rho-\rho_o) .
	\end{align*}
	
	\subsection{The Internal and Free Energy Densities}

The only remaining issue is to specify the internal energy density $e$ so that we can compute its  time derivative. To do so, it helps to  express the internal energy density, $e$, as a function of $T$ and $\rho$. These are more natural variables for the problem. To find $e = \hat{e}(T,\rho)$, first note that 
	\[
\frac{f}{T} = \frac{e}{T} - s,	
	\]
and, hence, it follows that 
	\[
d\left(\frac{f}{T} \right) = -\frac{e}{T^2}dT +\frac{1}{T} de -ds = -\frac{e}{T^2}dT + \frac{1}{T}  \left(\frac{\partial e}{\partial \rho}\right)_{s} d \rho .
	\]
Consequently,
	\begin{equation}
\left(\frac{\partial\left( \frac{f}{T} \right)}{\partial T}\right)_{\rho} = -\frac{e}{ T^2}  \implies e = -T^2 \left(\frac{\partial\left( \frac{f}{T} \right)}{\partial T}\right)_{\rho}.
	\label{eqn:free-to-internal}
	\end{equation}
We will begin with an expression of the form $f = f(T,\rho)$. In fact, for the PFC model, $f$ is usually modeled by an ideal gas law~\cite{elder07}. To gain some modeling flexibility, we will use a more general non-ideal gas law: 
	\[
f(T,\rho) = \rho k_BT \ln\left( \frac{\rho}{\rho_o}\right) -  k_BT\left(\rho -\rho_o\right) + \rho_o k_B T g\left(\frac{\rho-\rho_o}{\rho_o}\right) - \alpha \rho_o k_B T  \ln\left(\frac{T}{T_o}\right) - \beta\rho k_B T_o ,
	\]
where $k_B$ is Boltzmann's constant; $\rho_o$ is a reference density; $T_o>0$ is the reference temperature; $g$ is a polynomial that measures the deviation of the free energy density from the ideal gas model, and is typically non-negative; and, finally,  $\alpha$, and $\beta$ are positive, dimensionless constants. Clearly,
	\[
\left(\frac{\partial\left( \frac{f}{T} \right)}{\partial T}\right)_{\rho} = - \alpha \rho_o k_B \frac{1}{T} + \beta \rho k_B \frac{T_o}{T^2}.
	\]
It follows from \eqref{eqn:free-to-internal}, therefore, that the internal energy may be expressed as
	\[
e= \hat{e}(T,\rho) =  \alpha \rho_o k_B  T - \beta \rho k_B T_o,
	\]
which is linear in $T$ and $\rho$. Note that we use a hat over $e$, since this functional form is generally different from the functional form of $e$ in its natural thermodynamic coordinates $s$ and $\rho$.

For the PFC models, we define the latent heat as
	\[
\mathcal{L}(T) := e(T,\rho_L) - e(T,\rho_S) = \beta k_B T_o\left( \rho_S - \rho_L\right),
	\]
where $\rho_L$ is the (expected) spatially-uniform equilibrium density in the liquid phase at temperature $T$, and $\rho_S$ is the (expected) spatially-oscillatory field that characterizes the equilibrium solid phase at temperature $T$. These two density fields generally depend implicitly on the temperature. Otherwise, there is no direct dependence on the temperature for the latent heat. We will derive these fields shortly. In the applications that we will examine, the following inequality is valid: 
	\[
\rho_L = \overline{\rho_L} < \overline{\rho_S},	
	\]
where the overline represents the spatial average. In this case, the spatial average of the latent heat is a positive constant, since $\beta$ is assumed positive:
	\[
\overline{\mathcal{L}(T)} = \beta k_B T_o\left( \overline{\rho_S} - \overline{\rho_L}\right) > 0 .
	\]
As we see below, the average densities play an important role in the phase diagram and, therefore, the dynamics of the model.

	\subsection{The Full Model}

Finally, we can find the time derivative of the internal energy density:
	\[
\dot{e} = \alpha \rho_o k_B  \dot{T} - \beta \dot\rho k_B T_o.
	\]
With some other standard choices, namely,
	\[
M_T = M_{T,o} T^2, \quad M_\rho = M_{\rho,o}\rho ,	
	\]
where $M_{T,o}$ and $M_{\rho,o}$ are positive constants, the system becomes
	\begin{align*}
\alpha \rho_o k_B  \dot{T} - \beta  k_B T_o \dot\rho & = M_{T,o} \Delta T ,
	\\
\dot{\rho} & =  M_{\rho,o} \nabla\cdot \left( \rho\nabla \mu\right)
	\\
\mu & = k_B  \ln\left( \frac{\rho}{\rho_o}\right) + k_Bg'\left(\frac{\rho-\rho_o}{\rho_o}\right) - \beta k_B \frac{T_o}{T} + \kappa_{f,\rho,o} \mathcal{C} (\rho-\rho_o) .
	\end{align*}
We can now clearly identify the different terms in the chemical potential:	$k_B  \ln\left( \frac{\rho}{\rho_o}\right)$ corresponds to the ideal gas term, whereas $k_Bg'\left(\frac{\rho-\rho_o}{\rho_o}\right)$ takes care of the correction to the ideal gas law. The term $- \beta k_B \frac{T_o}{T}$ relates to the latent heat of fusion, and $\kappa_{f,\rho,o} \mathcal{C} (\rho-\rho_o)$ is concerned with the long-range interaction among particles. The boundary conditions are of local thermodynamic equilibrium (LTE) type, coupled with the entropy non-decreasing conditions
	\[
M_{T,o}T^2 \nabla \left(\frac{1}{T}\right) \cdot \bfn = -\frac{\beta_{T,\partial\Omega}}{T} , \quad - M_{\rho,o}\rho \nabla \mu\cdot \bfn  = \beta_{\rho,\partial\Omega} \mu \quad \mbox{on $\partial\Omega$}.
	\]
Recapitulating the entropy and energy densities, all of them as functions of $\rho$ and $T$, we have
	\begin{align*}
f = f(T,\rho) & = \rho k_BT \ln\left( \frac{\rho}{\rho_o}\right) -  k_BT\left(\rho -\rho_o\right) + \rho_o k_B T g \left(\frac{\rho-\rho_o}{\rho_o}\right) - \alpha \rho_o k_B T  \ln\left(\frac{T}{T_o}\right) - \beta\rho k_B T_o ,
	\\
e = \hat{e}(T,\rho) & = \alpha \rho_o k_B  T - \beta \rho k_B T_o ,
	\\
s = \hat{s}(T,\rho) & =    - \rho k_B \ln\left( \frac{\rho}{\rho_o}\right) +  k_B\left(\rho -\rho_o\right) -  \rho_o k_B  g\left(\frac{\rho-\rho_o}{\rho_o}\right) + \alpha \rho_o k_B\left(1+   \ln\left(\frac{T}{T_o}\right) \right) .
	\end{align*}

	\subsection{Non-dimensionalization}

By appropriate rescaling, we obtain the following non-dimensional version of the model:
	\begin{equation}
	\begin{aligned}
\alpha  \dot{T} - \beta \dot\rho  & = - M  \nabla\cdot\left( T^2 \nabla \left(\frac{1}{T}\right) \right) ,
	\\
\dot\rho & = \nabla\cdot \left( \rho \nabla \mu\right) ,
	\\
\mu & = \ln\left( \rho \right) + g'(\rho-1) -  \frac{\beta}{T} + \kappa\gamma(\rho-1) + \kappa\left(2 \Delta \rho  + \Delta^2 \rho\right) ,
	\end{aligned}
	\end{equation}
where $T \longrightarrow T/T_o$, $\rho\longrightarrow \rho/\rho_o$, and $M$, $\alpha$, $\beta$, $\varepsilon$, and $\kappa$, are positive, non-dimensional constants. The dimensionless constant $\gamma$ can be  positive, negative, or zero. The dimensionless boundary conditions are of LTE type, plus the following entropy producing conditions
	\[
-  T^2  \nabla \left(\frac{1}{T}\right) \cdot \bfn  = \frac{ \beta_{T,\partial\Omega}}{T} , \quad - \rho \nabla \mu\cdot \bfn  = \beta_{\rho,\partial\Omega} \mu \quad \mbox{on $\partial\Omega$},
	\]
where $\beta_{T,\partial\Omega}\ge 0$ and $\beta_{\rho,\partial\Omega}\ge0$. The dimensionless entropy is
	\[
S =  \int_{\Omega}\left\{ - \rho  \ln\left( \rho \right)  + \rho -1 - g(\rho-1)  + \alpha \left(1+  \ln\left(T\right) \right) - \frac{\kappa}{2}\left( \gamma(\rho-1)^2 - 2  | \nabla \rho|^2 + ( \Delta \rho)^2 \right) \right\}d\bfx ,
	\]
and the entropy production rate is
	\begin{equation}
\dot{S} =  \int_{\Omega}\left\{   \rho\left| \nabla \mu\right|^2  + M T^2 \left|\nabla \left(\frac{1}{T}\right) \right|^2   \right\}d\bfx + \int_{\partial\Omega} \left\{ \beta_{\rho,\partial\Omega} \mu^2  + \frac{\beta_{T,\partial\Omega}}{T^2}\right\} d\bfa \ge 0.
	\end{equation}

	\section{Free Energies and Phase Diagrams}
	\label{sec:phase-diagrams}
	
Let us examine the Helmholtz free energies of the liquid and solid phases, which inform the phase diagrams of the material. We say phase diagrams because each choice of the parameter set will give a different phase diagram. Understanding the free energy landscape will help us calibrate the melting temperature, $T_M$, as well as the equilibrium values of the fields $\rho_S$ and $\rho_L$. In particular, if we take $T_o = T_M$, at the melting temperature we should have $T=1.0$. We will show how to adjust parameters so this corresponds to the correct physical case. We will first recount the method to approximate the free energy analytically~\cite{provatas10}, and later we will explain how to get more accurate calculations numerically.

We will start by considering the non-dimensional free energy at the uniform, dimensionless temperature $T$:
	\begin{align*}
F[T,\rho] = \int_\Omega\bigg\{ T \rho \ln\left( \rho \right) -  T\left(\rho - 1\right) + g(\rho-1) - \alpha  T  \ln\left(T\right) - \beta\rho   + \frac{\kappa T}{2}\left( \gamma(\rho - 1)^2 - 2  | \nabla \rho|^2 + ( \Delta \rho)^2 \right) \bigg\}d\bfx .
	\end{align*}
The phase diagram can be constructed by minimizing the free energy, or, equivalently, maximizing the entropy, as long as the temperature is uniform, which we assume in this section. Now, we will make a small deviation approximation. Suppose that
	\[
\rho = 1+ \psi , \quad |\psi| \ll 1.	
	\]
For simplicity, let us assume for now that the ideal gas deviation term, $g$, is identically zero. (We will take up the more general case in future papers.)	Then, using Taylor's Theorem,
	\begin{equation}
 T\rho\ln(\rho) - T(\rho-1) -\beta \rho + \frac{\kappa\gamma T}{2}(\rho-1)^2  =  \frac{\lambda T}{2} \psi^2 - \frac{T}{6}\psi^3 + \frac{T}{12} \psi^4 - \beta(\psi+1) + \mathcal{O}(\psi^5),
	\end{equation}
where
	\[
\lambda := 1 + \gamma\kappa.
	\]
Thus, the approximate free energy is
	\begin{equation}
\mathcal{F}[T,\psi] = \int_\Omega\left\{ \frac{\lambda T}{2} \psi^2 - \frac{T}{6}\psi^3 + \frac{ T}{12}\psi^4 - \beta(\psi+1) - \alpha T\log(T) +\frac{\kappa T}{2}\left( - 2  | \nabla\psi|^2 + ( \Delta\psi)^2 \right) \right\}d\bfx .
	\label{eqn:poly-free-energy}
	\end{equation}
	
We will construct a 2D phase diagram using the procedure outlined in~\cite{provatas10}. The 3D version can be done in an analogous way. See, for example, \cite{provatas10}. In the crystalline phase, we observe through computations that the density field has a spatially oscillatory ``equilibrium" solution that is, up to rotations, approximately of the form
	\begin{equation}
\psi(x,y) \approx \chi(x,y) := A +B \left(\cos\left(\frac{2 \pi x}{p}\right) \cos\left(\frac{2 \pi y}{\sqrt{3} p}\right) + \frac{1}{2} \cos\left(\frac{4 \pi y}{\sqrt{3} p}\right)\right),
	\label{eqn:solid-ansatz}
	\end{equation}
where $A$, $B$, and $p>0$ are parameters to be determined. The approximation $\chi$ is sometimes called the crystal ansatz. Observe that
	\[
\overline{\chi} = A,	
	\]
that is, $A$ is the average density and $B$ is the amplitude of oscillations. This solution has hexagonal symmetry. Its peaks form a hexagonal Bravais lattice. (See Figure~\ref{fig:solid-sim}.) Next, fixing $\alpha$, $\beta$, $\kappa$, and $\lambda$, we define
	\begin{align*}
& \hspace{-0.1in} G(A,B,p,T) 
	\\
&:= \frac{2}{\sqrt{3}p^2}\int_0^{\frac{p\sqrt{3}}{2}}\int_0^p \left\{ \frac{\lambda T}{2} \chi^2 - \frac{T}{6}\chi^3 + \frac{T}{12}\chi^4 - \beta(\chi+1) - \alpha T\log(T)  +\frac{\kappa T}{2}  \left( -2  |\nabla\chi|^2 +\left(  \Delta \chi\right)^2 \right) \right\}dxdy, 
	\end{align*} 
which represents the free energy at a constant temperature $T$ evaluated at the approximate solution $\chi$, averaged over the crystal's unit cell, that is, the smallest repeat unit of $\chi$.

Taking the derivative with respect to $p$, setting this equal to zero, and solving for $p$, we have
	\[
\frac{\partial G}{\partial p} = \frac{4 B^2 \kappa (3 p^2 \pi^2 - 16 \pi^4) T}{ 3 p^5} = 0 \quad \implies \quad p = p_{\rm eq} := \frac{4\pi}{\sqrt{3}} .
	\]
This gives
	\begin{align*}
G(A,B,p_{\rm eq},T) & = -\beta(1 + A) + \frac{T}{1536} \bigg(-256 A^3 - 48 B^2 (B + 6 \kappa - 6 \lambda) + 128 A^4   + 45 B^4   
	\\
& \quad +   96 A B^2 (-3 + B  ) + 96 A^2 (8 \lambda + 3 B^2  )\bigg) - \alpha T\log(T) .
	\end{align*}
	
Next, taking the derivative with respect to $B$, we have
	\[
\frac{\partial }{\partial B} G(A,B,p_{\rm eq},T) = \frac{3 B T}{128}  \left(-4 (4 A + B + 4 \kappa - 4 \lambda) + (16 A^2 + 8 A B + 5 B^2)  \right). 
	\]
This cubic equation (with respect to $B$) has the solutions $B = 0$, which represents the liquid state, and the two crystalline solutions
	\begin{equation}
	\label{eqn:amplitude-approx}
B = B_{\pm} := \frac{2 - 4 A   \pm  2\sqrt{ 1 + 16 A   + 20 ( \kappa - \lambda) - 16 A^2}}{5  }.
	\end{equation}
We will take $B = B_+$, since, it turns out, using $B = B_-$ results in a higher free energy for the crystal ansatz. 
	
	\begin{figure}
	\begin{center}
\includegraphics[width = 4.0in]{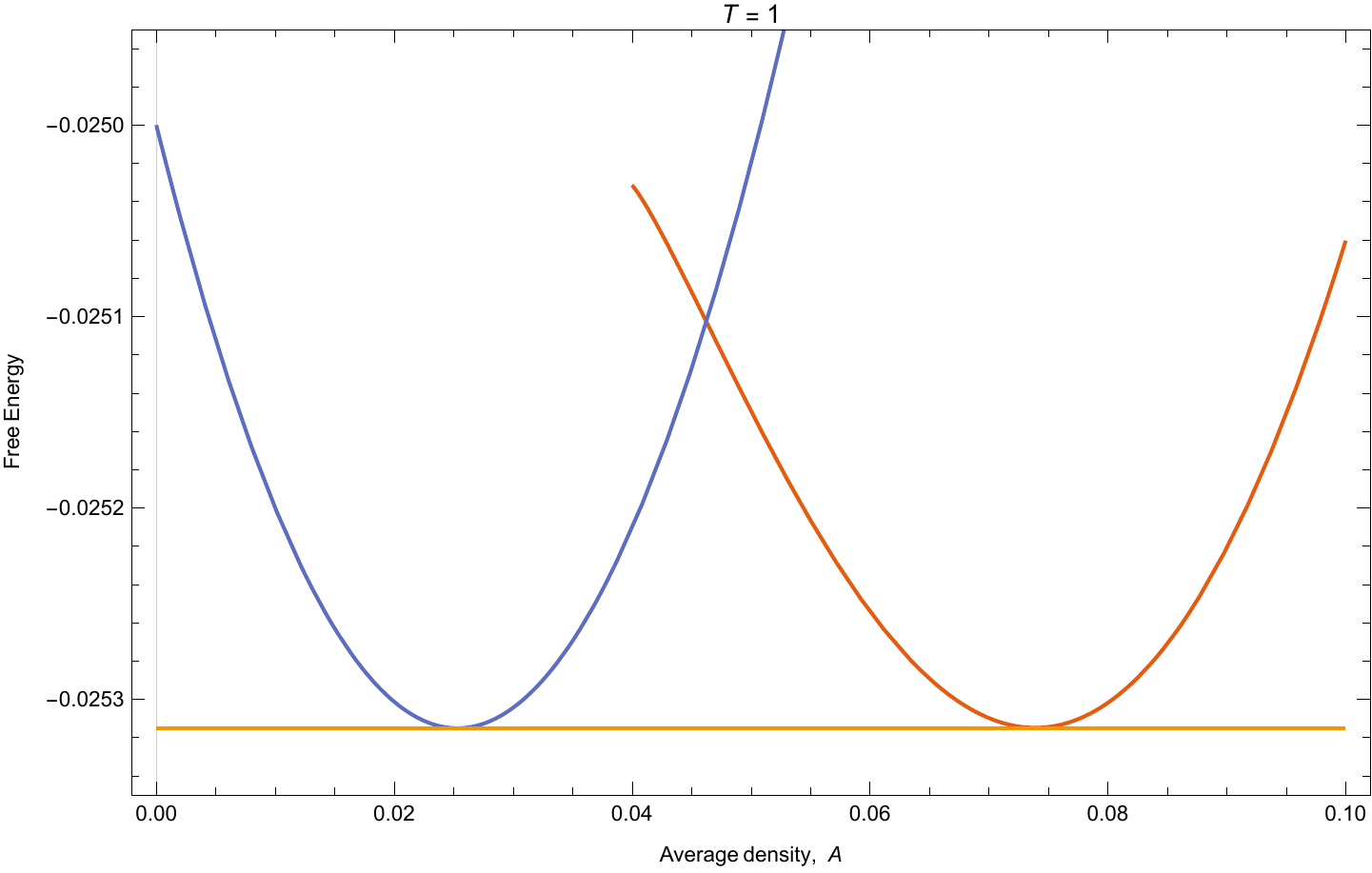}
	\end{center}
\caption{Plots of the free energies (per unit cell) $\mathcal{F}_{S}$ (red) and $\mathcal{F}_{L}$ (blue) for the parameters $T = 1.0$, $\lambda = 1.0$, $\kappa = 0.919285$, $\beta = 0.025$, and $\alpha = 0.1$. At the melting temperature, $T=1.0$, the minimum values of $\mathcal{F}_{S}$ and $\mathcal{F}_{L}$ are equal. The orange line is $-0.025315$.}
	\label{fig:melting}
	\end{figure}

	\begin{figure}
	\begin{center}
\includegraphics[width = 3.0in]{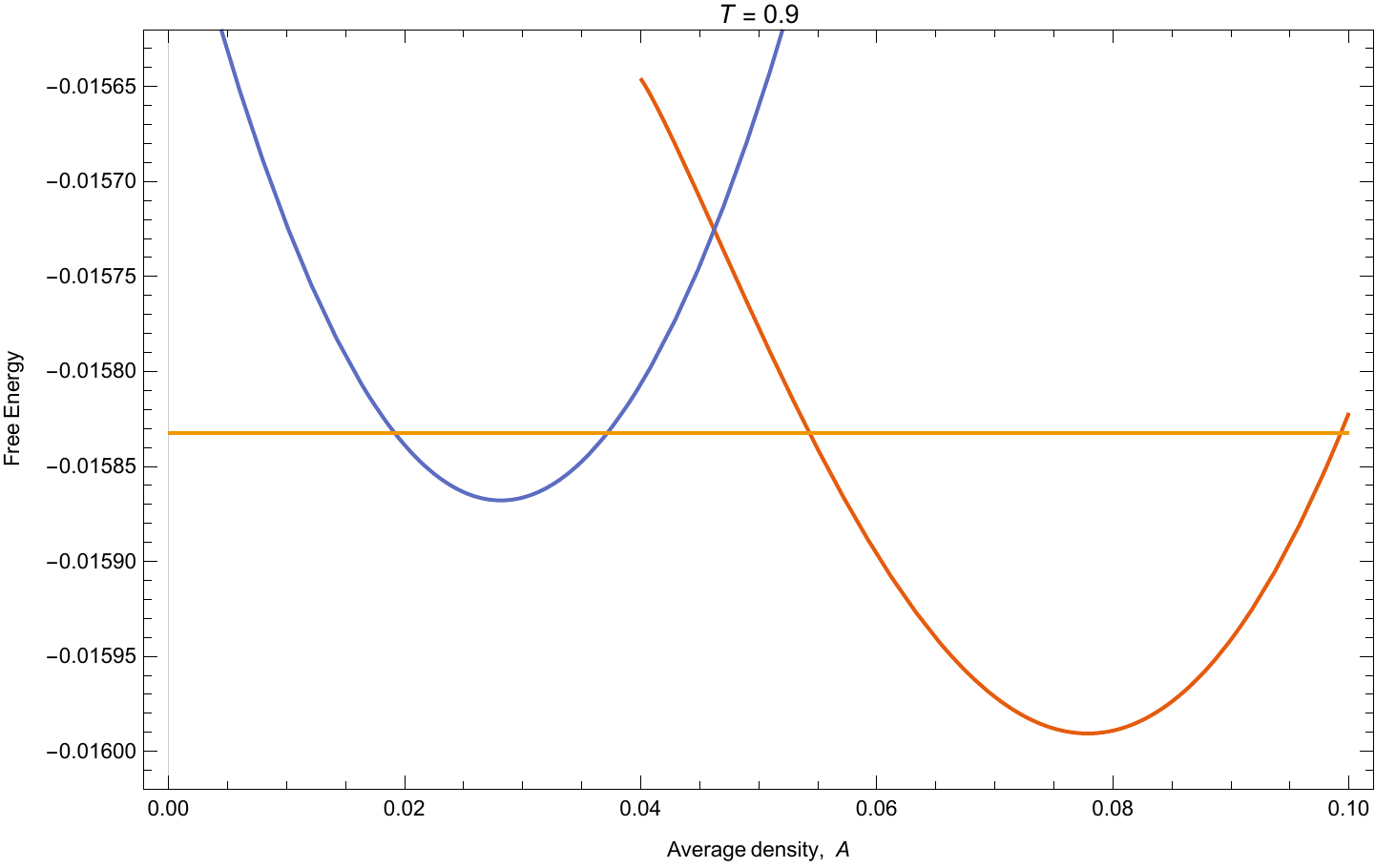}
\hspace{0.25in}
\includegraphics[width = 3.0in]{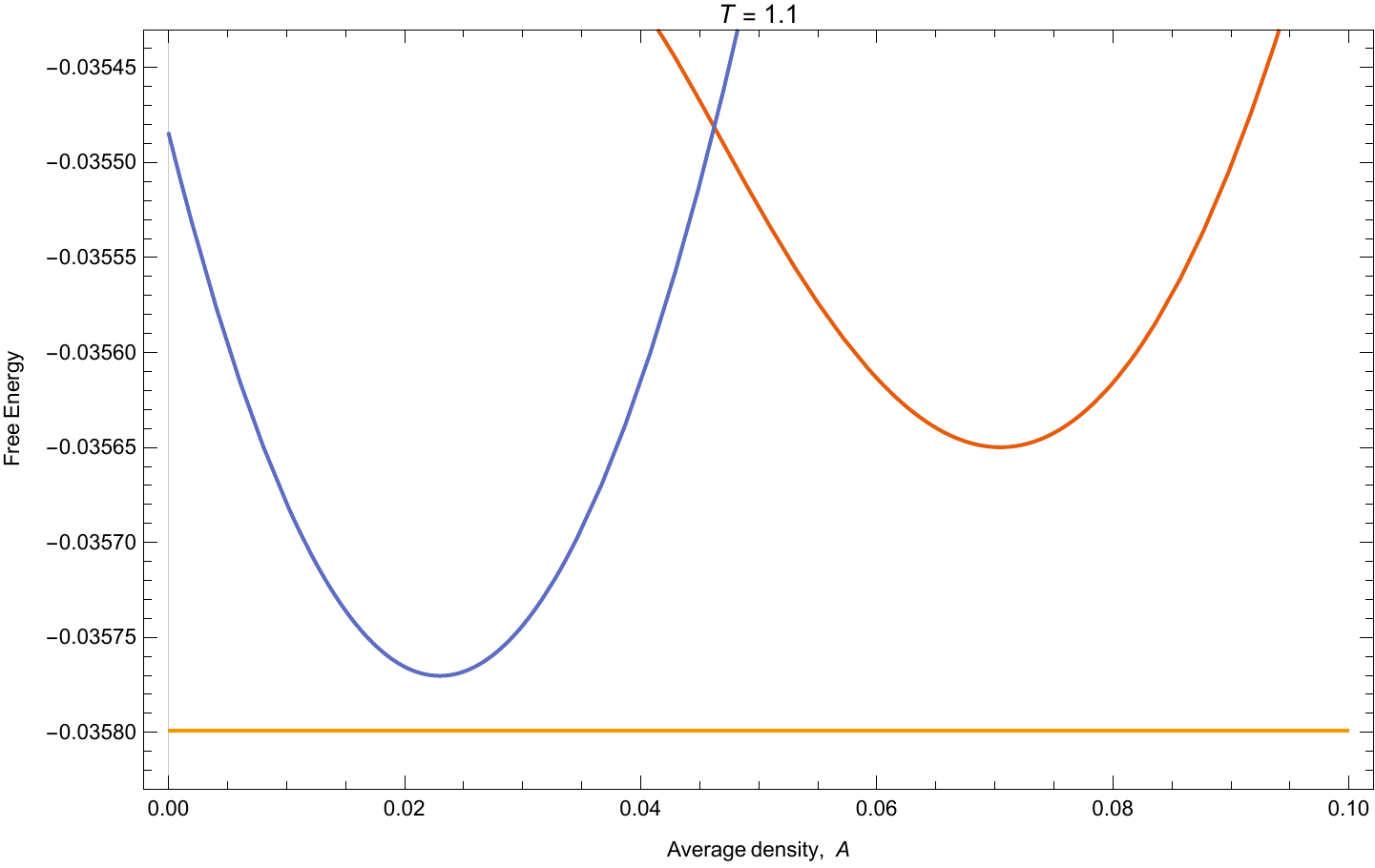}
	\end{center}
\caption{Plots of the free energies (per unit cell) $\mathcal{F}_{S}$ and $\mathcal{F}_{L}$ for the parameters $T=0.9$ (left), $T = 1.1$ (right), and $\lambda = 1.0$, $\kappa = 0.919285$, $\beta = 0.025$, and $\alpha = 0.1$. When the temperature is above (below) the melting temperature, $T=1.0$, the liquid phase has lower (higher) free energy. The orange line is $-0.025315 -\alpha T \log(T)$.}
	\label{fig:colder-warmer}
	\end{figure}	
	
Now, the liquid free energy per unit cell can be obtained simply by setting $B = 0$. Thus we define
	\begin{equation}
\mathcal{F}_{L}(A,T) = G(A,0,p,T) = \frac{\lambda T}{2} A^2 - \frac{T}{6} A^3 + \frac{ T}{12} A^4 - \beta(A+1) - \alpha T\log(T).
	\label{eqn:exact-liquid}
	\end{equation}
This expression is exact, assuming that \eqref{eqn:poly-free-energy} is the exact free energy expression. The crystal phase free energy per unit cell is approximated as 
	\[
\mathcal{F}_{S}(A,T) = G(A,B_+,p_{\rm eq},T).
	\]


Now, the melting temperature, $T=1.0$, should be, by definition, the temperature at which the minimum values of $\mathcal{F}_{S}(A,T)$ and $\mathcal{F}_{L}(A,T)$ are equal. Thus, for a given parameter set $\lambda$, $\kappa$, $\beta$, $\alpha$,
	\begin{equation}
\mathcal{F}_{S}(A_{S},1) = 	\mathcal{F}_{L}(A_{L},1),
	\label{eqn:free-energies-T=1}
	\end{equation}
where
	\[
A_{L} := \operatorname{argmin}_{A} \mathcal{F}_{L}(A,1) \quad \mbox{and}\quad A_{S} := \operatorname{argmin}_{A} \mathcal{F}_{S}(A,1)	.
	\]
The parameters can be carefully adjusted to make equality in \eqref{eqn:free-energies-T=1} happen. We will show a couple of examples.

In Figure~\ref{fig:melting}, the free energies (per unit cell) $\mathcal{F}_{S}$ (red) and $\mathcal{F}_{L}$ (blue) are plotted for the parameters $T = 1.0$, $\lambda = 1.0$, $\kappa = 0.919285$, $\beta = 0.025$, and $\alpha = 0.1$. At the melting temperature, $T=1.0$, the minimum values of $\mathcal{F}_{S}$ and $\mathcal{F}_{L}$ are equal, as shown. If the temperature is below the melting temperature, the free energy of the solid phase should be lower. This case is shown in Figure~\ref{fig:colder-warmer}(left). Conversely, if the temperature is above the melting temperature, the free energy of the liquid phase should be lower than that of the solid phase. This case is shown in Figure~\ref{fig:colder-warmer}(right). Thus the model is flexible enough so that (i) the melting temperature can be calibrated so that \eqref{eqn:free-energies-T=1} holds at $T=1.0$; (ii) if $T<1.0$, the solid phase has lower free energy than the liquid phase; and (iii) if $T >1.0$, the solid phase has higher free energy than the liquid phase. Our latent heat term is simple in this model; it is linear in $\rho$ ($-\beta\rho$). But it is, seemingly, sophisticated enough to capture the basic physics near the melting temperature. 

To compute phase diagrams, one uses the Maxwell common-tangents construction. The common tangent approach is appropriate when the number density of particles is  a conserved quantity, as is typical with the PFC modeling framework. The Maxwell construction method is described in detail in~\cite{DeHoff1993} and is understood widely in the Physics literature, so we will not describe it here. See also~\cite{provatas10}.

	\begin{figure}
	\begin{center}
\includegraphics[width = 4.0in]{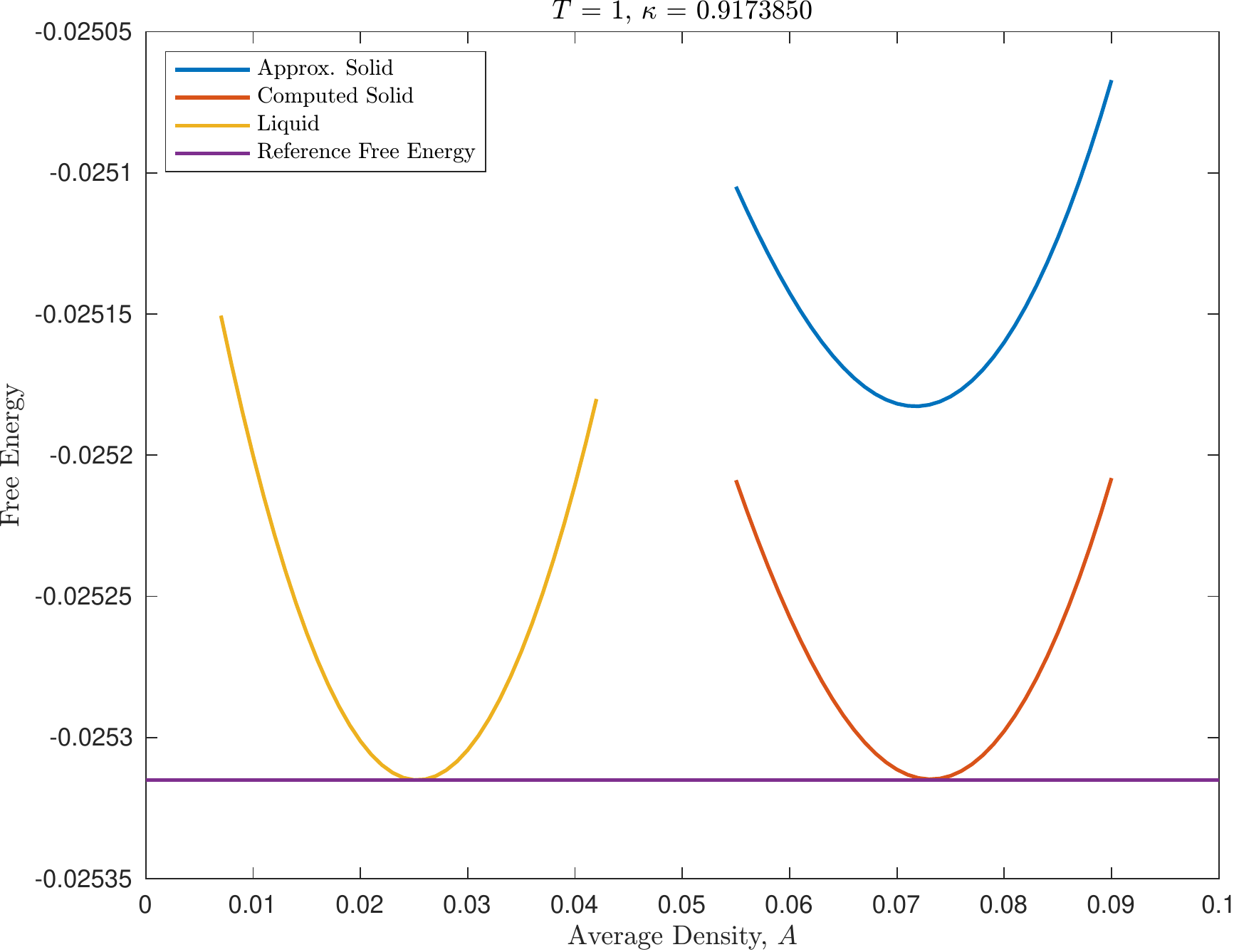}
	\end{center}
\caption{Plots of the free energies (per unit cell) $\mathcal{F}^{\rm E}_{S}$ (red), $\mathcal{F}_{S}$ (blue), and $\mathcal{F}_{L}$ (yellow) for the parameters $T = 1.0$, $\lambda = 1.0$, $\kappa = 0.917385$, $\beta = 0.025$, and $\alpha = 0.1$. At the (recalibrated) melting temperature, $T=1.0$, the minimum values of $\mathcal{F}^{\rm E}_{S}$ and $\mathcal{F}_{L}$ are equal. The purple line is $-0.025315$. Notice that, as expected, the approximate solution obtained by the ansatz~\eqref{eqn:solid-ansatz} yields a higher free energy. We adjusted $\kappa$ lower (relative to the value in Figure~\ref{fig:melting}) in order to recalibrate the parameters so that $T=1.0$ was the ``exact" melting temperature.}
	\label{fig:melting-exact-1}
	\end{figure}

	\begin{figure}
	\begin{center}
\includegraphics[width = 3.0in]{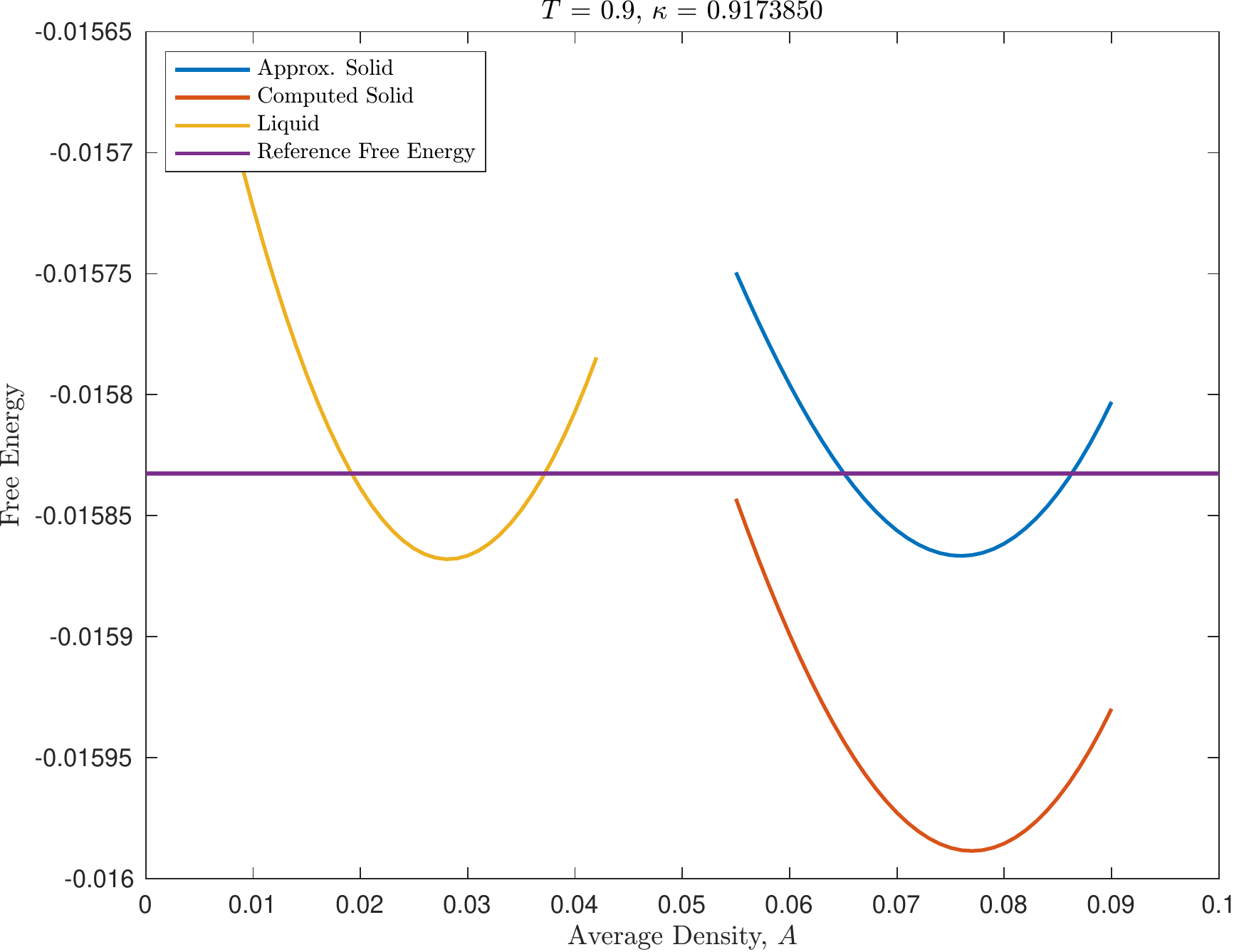}
\hspace{0.25in}
\includegraphics[width = 3.0in]{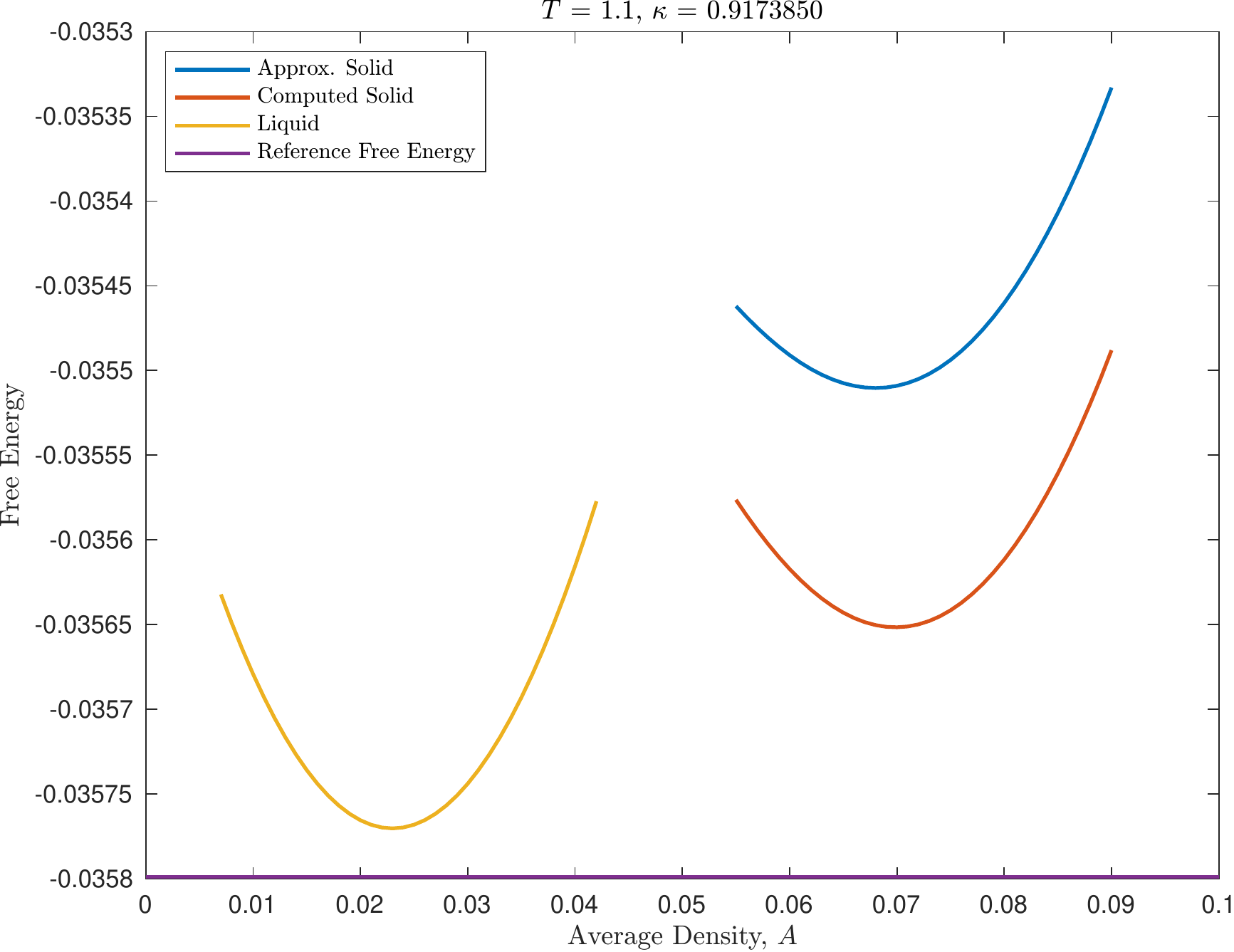}
	\end{center}
\caption{Plots of the free energies (per unit cell) $\mathcal{F}^{\rm E}_{S}$ (red), $\mathcal{F}_{S}$ (blue), and $\mathcal{F}_{L}$ (yellow) for the parameters $T = 0.9$ (left), $T = 1.1$ (right), and $\lambda = 1.0$, $\kappa = 0.917385$, $\beta = 0.025$, and $\alpha = 0.1$. The uniform temperature is below (above) the recalibrated melting temperature, and the minimum value of $\mathcal{F}^{\rm E}_{S}$ is below (above) that of $\mathcal{F}_{L}$. The purple line is $-0.025315-\alpha T\log(T)$.}
	\label{fig:colder-warmer-exact-1}
	\end{figure}

We have computed the free energy landscape using the crystal solution ansatz~\eqref{eqn:solid-ansatz}, which is only an approximation of the solid phase minimizer. In particular, the approximated free energy is expected to be too large. How big of an error is incurred by using the approximation? To find out, let us describe a more accurate method for computing the free energy landscape.  In particular, we need a more accurate representation of the solid phase free energy. For illustration purposes, let us suppose that the free energy, $\mathcal{F}$, as given in  \eqref{eqn:poly-free-energy}, is  the ``exact" free energy. The liquid free energy per unit cell \eqref{eqn:exact-liquid} is the same as before, because no approximation was used to obtain it. For the crystalline (solid) state, instead of using the ansatz~\eqref{eqn:solid-ansatz}, we will use numerical calculations to find the exact energy (per unit cell)
	\[
\mathcal{F}^{\rm E}_{S}(A,T),
	\]
where $A := \overline{\rho -1}$, as a function of $A$, at a uniform fixed temperature, $T$, and a given set of parameters $\lambda$, $\kappa$, $\beta$, $\alpha$. The procedure for calculating $\mathcal{F}^{\rm E}_{S}$ is outlined as follows: fixing the parameters $p$, $A$, $T$, $\lambda$, $\kappa$, $\beta$, and $\alpha$, we solve the system
	\begin{equation}
	\label{eqn:PFC-trad}
\dot{\rho} = \Delta \mu , \quad \mu = \delta_\rho \mathcal{F}[T,\rho], \quad \mbox{on} \quad \Omega = \Omega_p := [0,p]\times \left[0, \frac{\sqrt{3}}{2} p\right]
	\end{equation}
subject to periodic boundary conditions, with the initial conditions given by the crystal ansatz:
	\[
\rho(x,y,t=0) = 1 + A +B_+ \left(\cos\left(\frac{2 \pi x}{p}\right) \cos\left(\frac{2 \pi y}{\sqrt{3} p}\right) + \frac{1}{2} \cos\left(\frac{4 \pi y}{\sqrt{3} p}\right)\right),
	\]
where $B_+$ is a given in \eqref{eqn:amplitude-approx}. Clearly,
	\[
\overline{\rho( \, \cdot \, , 0) -1} = A.	
	\]
We solve \eqref{eqn:PFC-trad} to steady state (equilibrium) and compute the energy of the solution on the unit cell, obtaining the equilibrium solutions
	\[
\rho_\infty \equiv \rho(\, \cdot \, , t=\infty) \quad \mbox{on} \quad   [0,p]\times \left[0, \frac{\sqrt{3}}{2} p\right]
	\]
and
	\begin{align*}
F_\infty(A,T,p) &:= \int_0^p\int_0^{\sqrt{3} p/2}\bigg\{ \frac{\lambda T}{2} (\rho_\infty-1)^2 - \frac{T}{6}(\rho_\infty-1)^3 + \frac{ T}{12}(\rho_\infty-1)^4 - \beta \rho_\infty 
	\\
& \quad - \alpha T\log(T) +\frac{\kappa T}{2}\left( - 2  | \nabla\rho_\infty|^2 + ( \Delta\rho_\infty)^2 \right) \bigg\}dy dx .
	\end{align*}
Finally,
	\[
\mathcal{F}^{\rm E}_{S}(A,T) := F_\infty(A,T,p^{\rm E}_{\rm eq}), \quad p^{\rm E}_{\rm eq} := \operatorname{argmin}_p F_\infty(A,T,p).
	\]
We use a pseudo-spectral, stabilized implicit-explicit (IMEX) method to perform the equilibrium field calculations. The spectrally accurate trapezoidal rule is used to compute the energies. The minimization problem in $p$ is solved using a derivative-free method~\cite{brent73} over the interval $[p_{\rm eq}- 1,p_{\rm eq}+1]$, where $p_{\rm eq} = \frac{4\pi}{\sqrt{3}}$.

The results of our improved free energy computations are shown in Figures~\ref{fig:melting-exact-1}, $T=1.0$; \ref{fig:colder-warmer-exact-1}(left), $T=1.1$; and \ref{fig:colder-warmer-exact-1}(right), $T=0.9$. Notice that the approximate solution yields a larger free energy than the true minimizer, as expected. To recalibrate the model so that $T = 1.0$ occurs when $\mathcal{F}^{\rm E}_{S}(A_s^E,T) = \mathcal{F}_{L}(A_L,T)$, we need only to adjust the value of $\kappa$ lower. The approximate free energy landscape computed using the ansatz is still quite useful, since the resulting free energies are reasonably good ballpark estimates that can help tune the parameters. The more accurate method can then refine and recalibrate the parameters after the ballpark estimates are obtained.

	\section{Numerical Solution of the Model}
	\label{sec:numerics}
	
To conclude this paper, let us perform a simple computation to give just a small sample of what this model can do. We use the polynomial free energy~\eqref{eqn:poly-free-energy} as our ``exact" free energy;
the associated entropy is 
	\[
\mathcal{S} =  \int_{\Omega}\left\{ - \left( \frac{\lambda }{2} \psi^2 - \frac{1}{6}\psi^3 + \frac{1}{12}\psi^4 \right)  + \alpha \left(1+  \ln\left(T\right) \right) - \frac{\kappa}{2}\left(  - 2  | \nabla \psi|^2 + ( \Delta \psi)^2 \right) \right\}d\bfx ;
	\]
and the associated internal energy becomes  
	\[
\mathcal{E} = \int_\Omega \left\{ \alpha T - \beta(\psi+1) \right\} d\bfx ,	
	\]
where we have made use of the change of variable
	\[
\psi := \rho - 1.	
	\]
The simplified evolution equations are
	\begin{equation}
	\begin{aligned}
\alpha  \dot{T} - \beta \dot\psi  & = M  \Delta T ,
	\\
\dot\psi & = \Delta \mu  ,
	\\
\mu & = \frac{1}{T}\delta_\psi\mathcal{F} =  \lambda\psi -\frac{1}{2}\psi^2 +\frac{1}{3}\psi^3  -  \frac{\beta}{T} + \kappa\left(2 \Delta \psi  + \Delta^2 \psi\right) ,
	\end{aligned}
	\end{equation}
and we use periodic boundary conditions for simplicity. We leave it to the reader to show that this system is still entropy producing, locally and globally. Mass is conserved in our simulation, and therefore, this simplified setup disallows freezing in the usual sense. This is because the solid and liquid phases have distinct average densities. If we start out with the equilibrium values of the solid and liquid states and then drop the temperature below the melting temperature, the solid may not grow much, if at all. We would have to add mass to the system in order for the solid state to grow, since the solid state has a higher average density. We will save such sophisticated simulations for a future paper, where we will explore the model further.

\begin{figure}
	\begin{center}
\includegraphics[width = 3.0in]{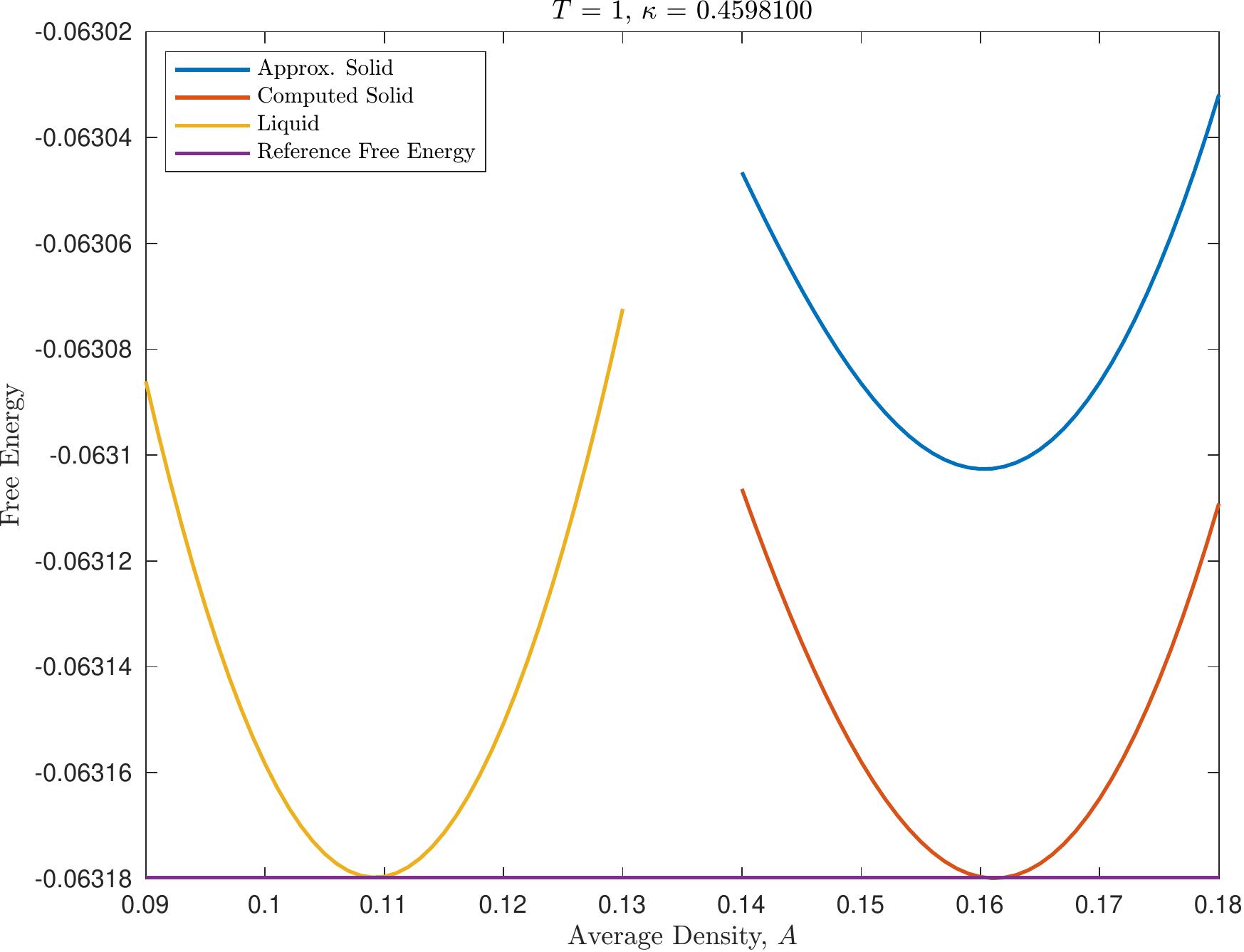}
\includegraphics[width = 3.0in]{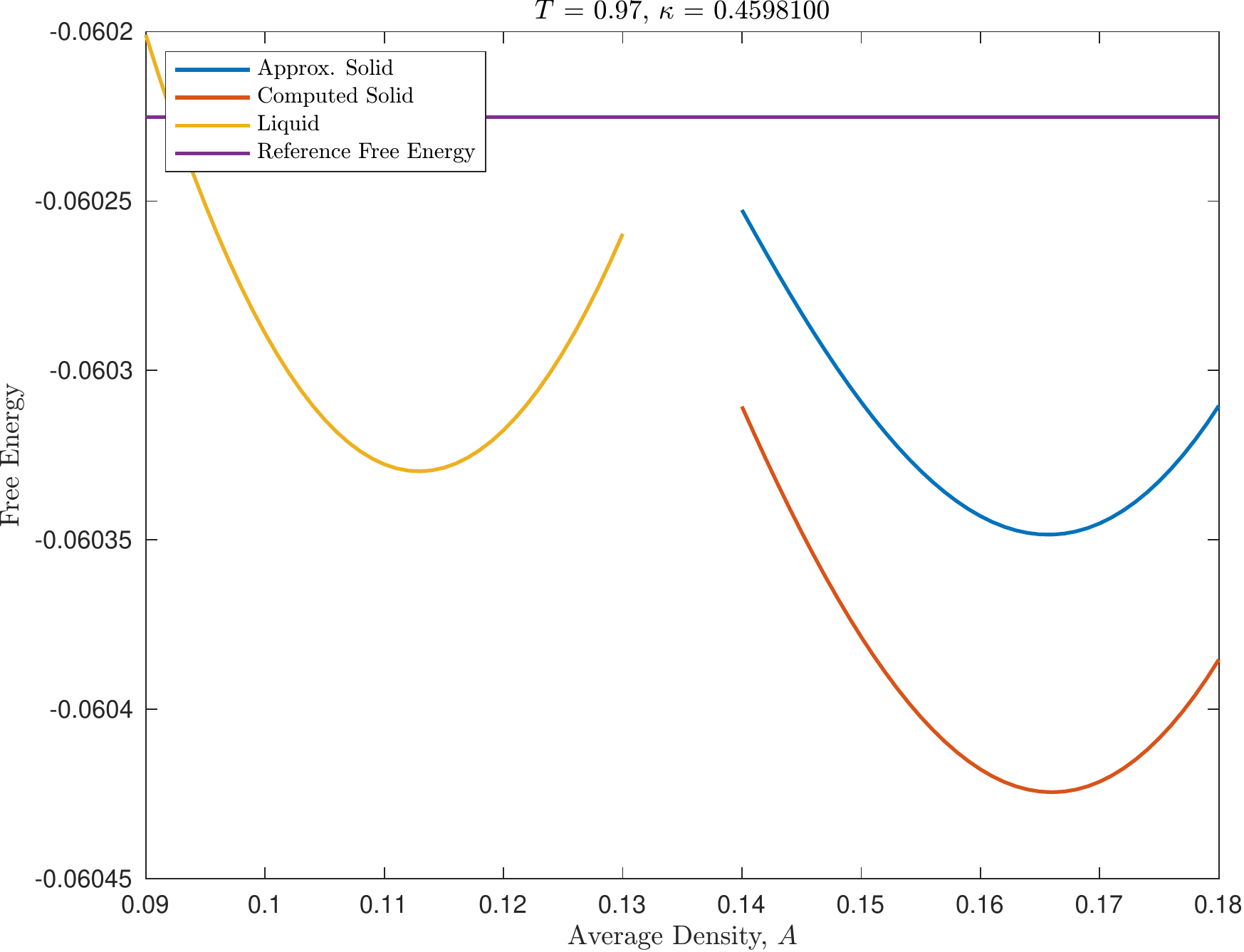}
	\end{center}
\caption{Plots of the free energies (per unit cell) $\mathcal{F}^{\rm E}_{S}$ (red), $\mathcal{F}_{S}$ (blue), and $\mathcal{F}_{L}$ (yellow) for the parameters $T = 1.0$ (left), $T = 0.97$ (right), and $\lambda = 0.6$, $\kappa = 0.459810$, $\beta = 0.06$, and $\alpha = 0.1$. The uniform temperature is equal to (below) the recalibrated melting temperature, and the minimum value of $\mathcal{F}^{\rm E}_{S}$ is equal to (below) that of $\mathcal{F}_{L}$. The purple line is $-0.0631798-\alpha T\log(T)$.}
	\label{fig:melting-exact-2}
	\end{figure}

\begin{figure}
	\begin{center}
$t = 5.0$
	\\
\includegraphics[width = 2.6in]{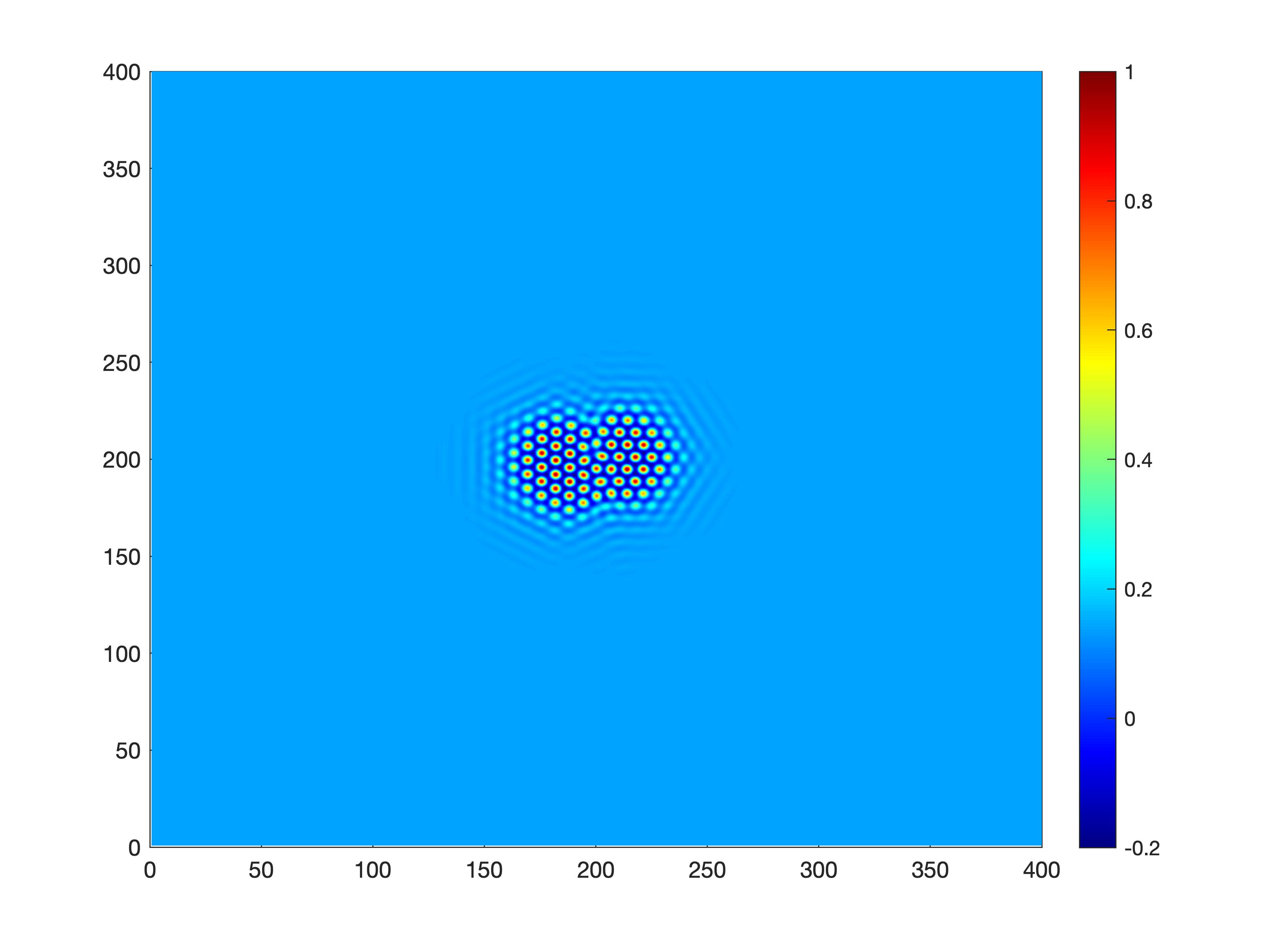}
\includegraphics[width = 2.6in]{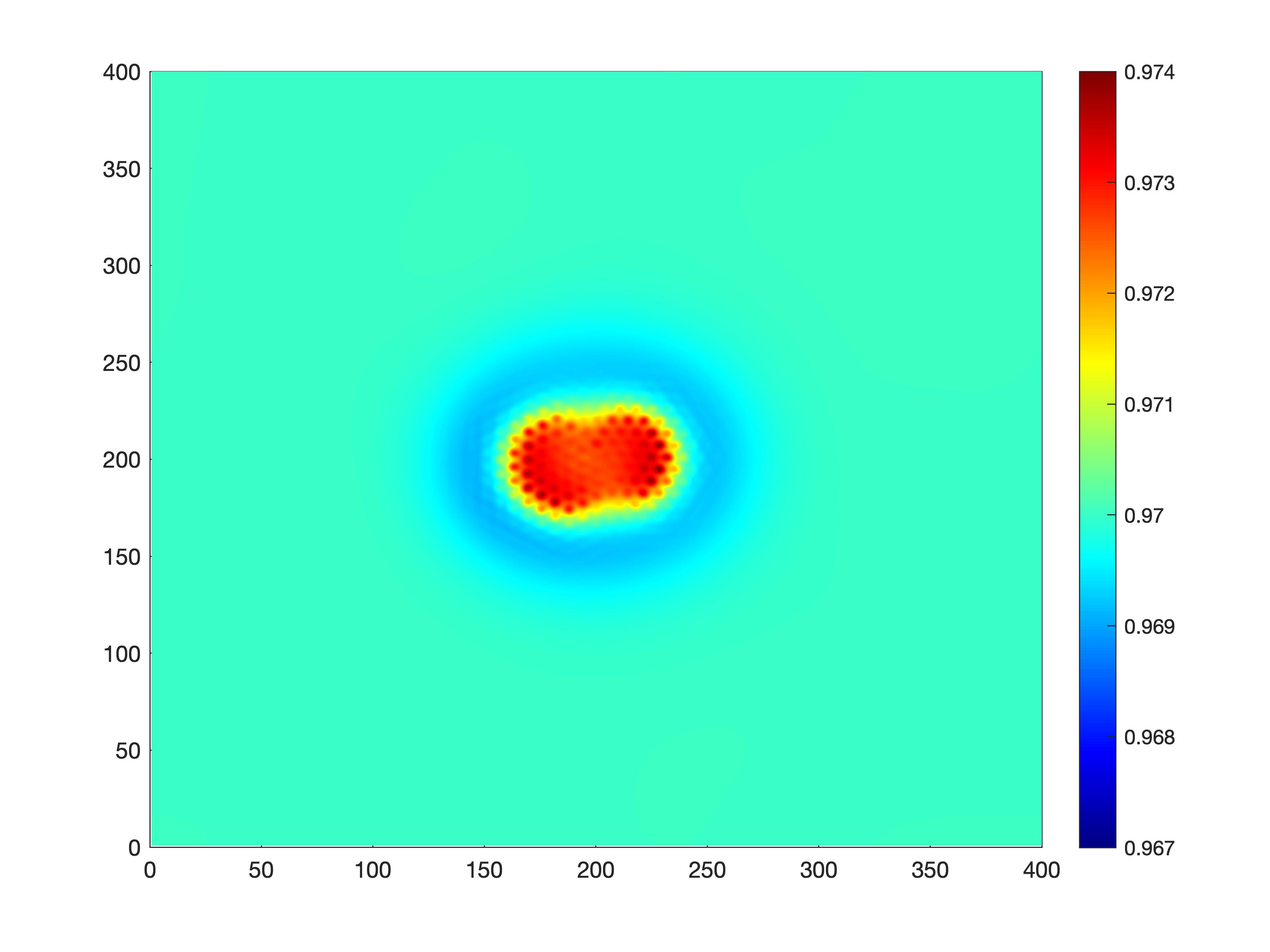}
	\\
$t = 10$
	\\
\includegraphics[width = 2.6in]{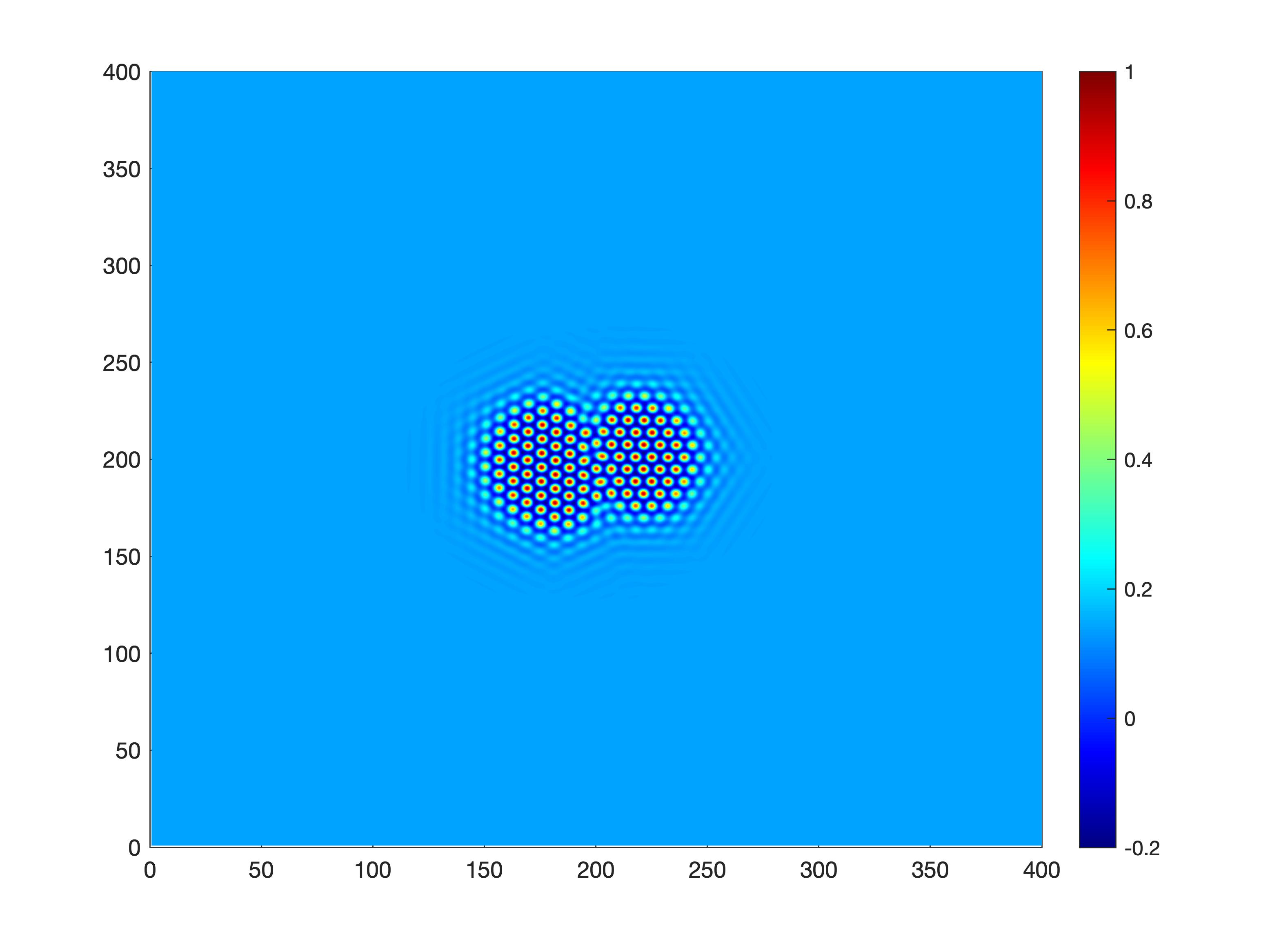}
\includegraphics[width = 2.6in]{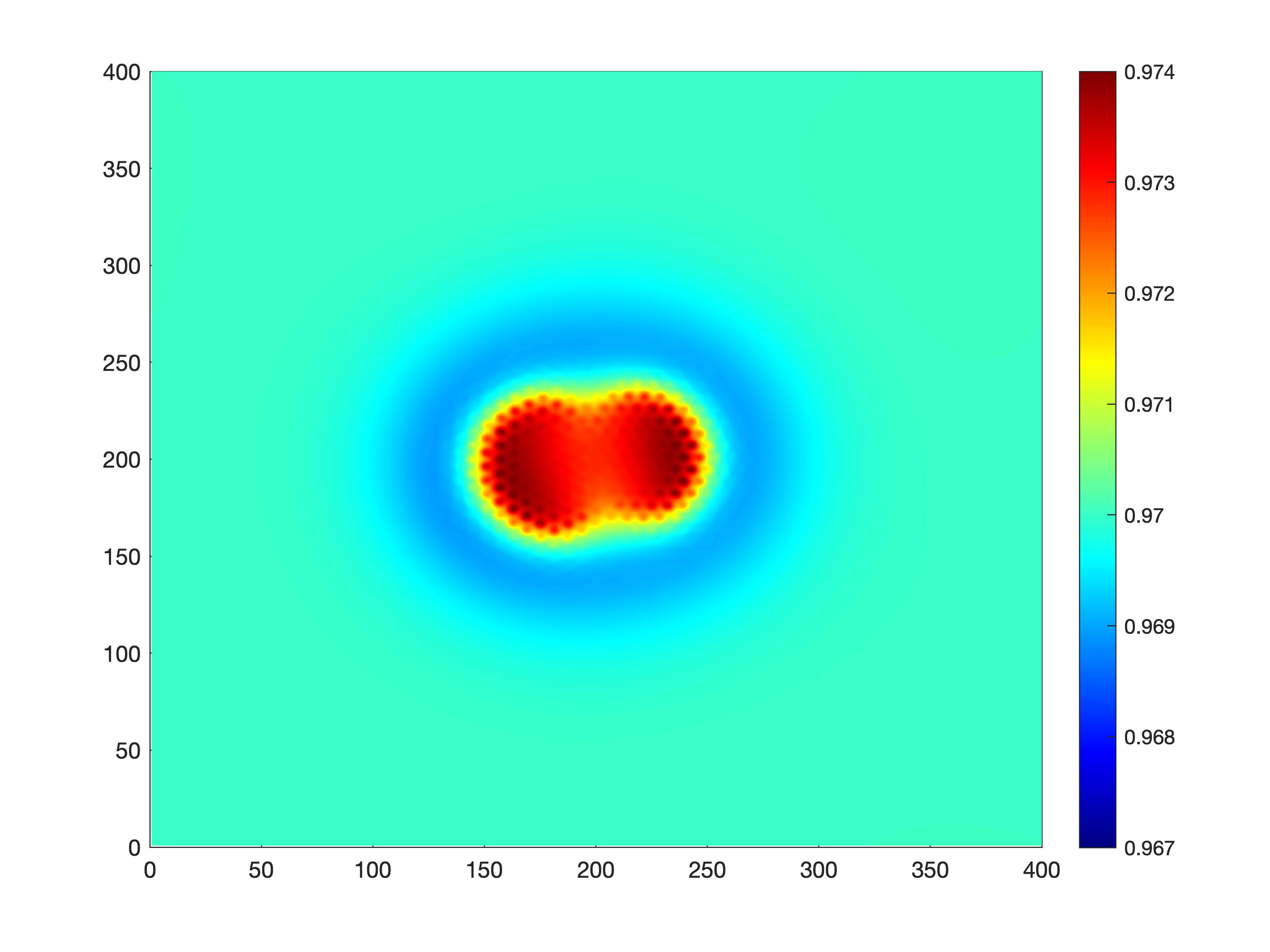}
	\\
$t = 50$
	\\
\includegraphics[width = 2.6in]{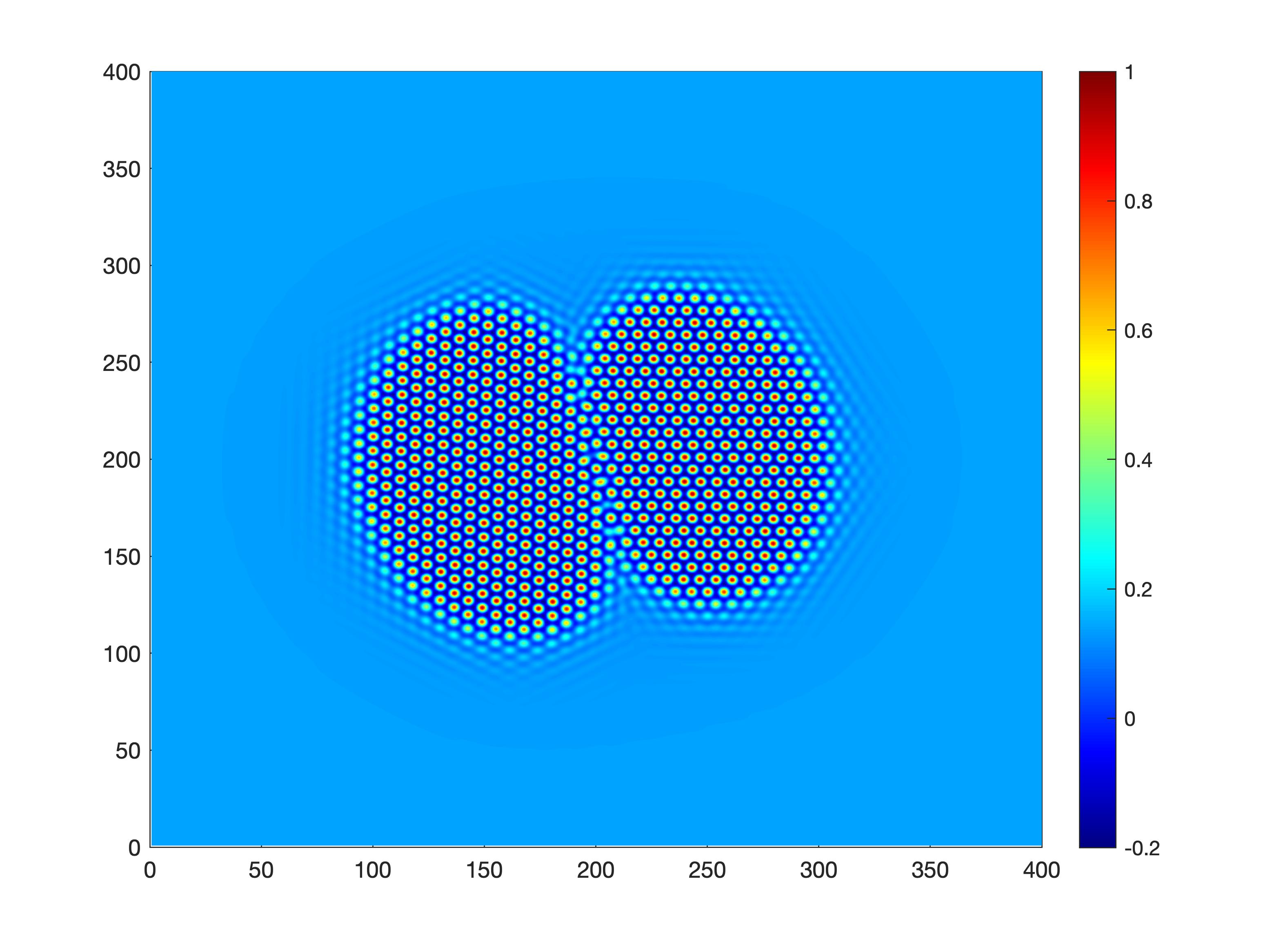}
\includegraphics[width = 2.6in]{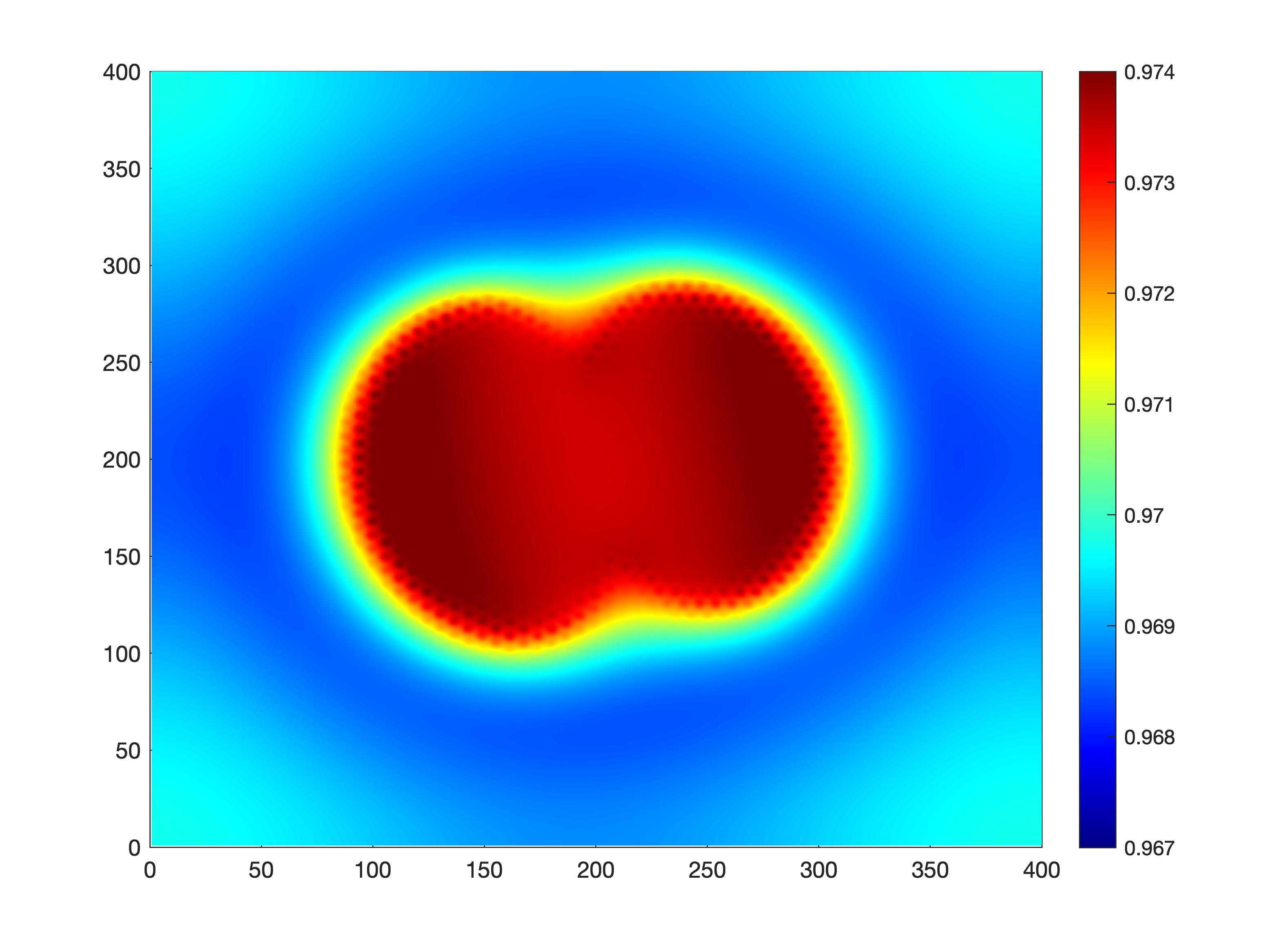}
	\\
$t = 100$
	\\
\includegraphics[width = 2.6in]{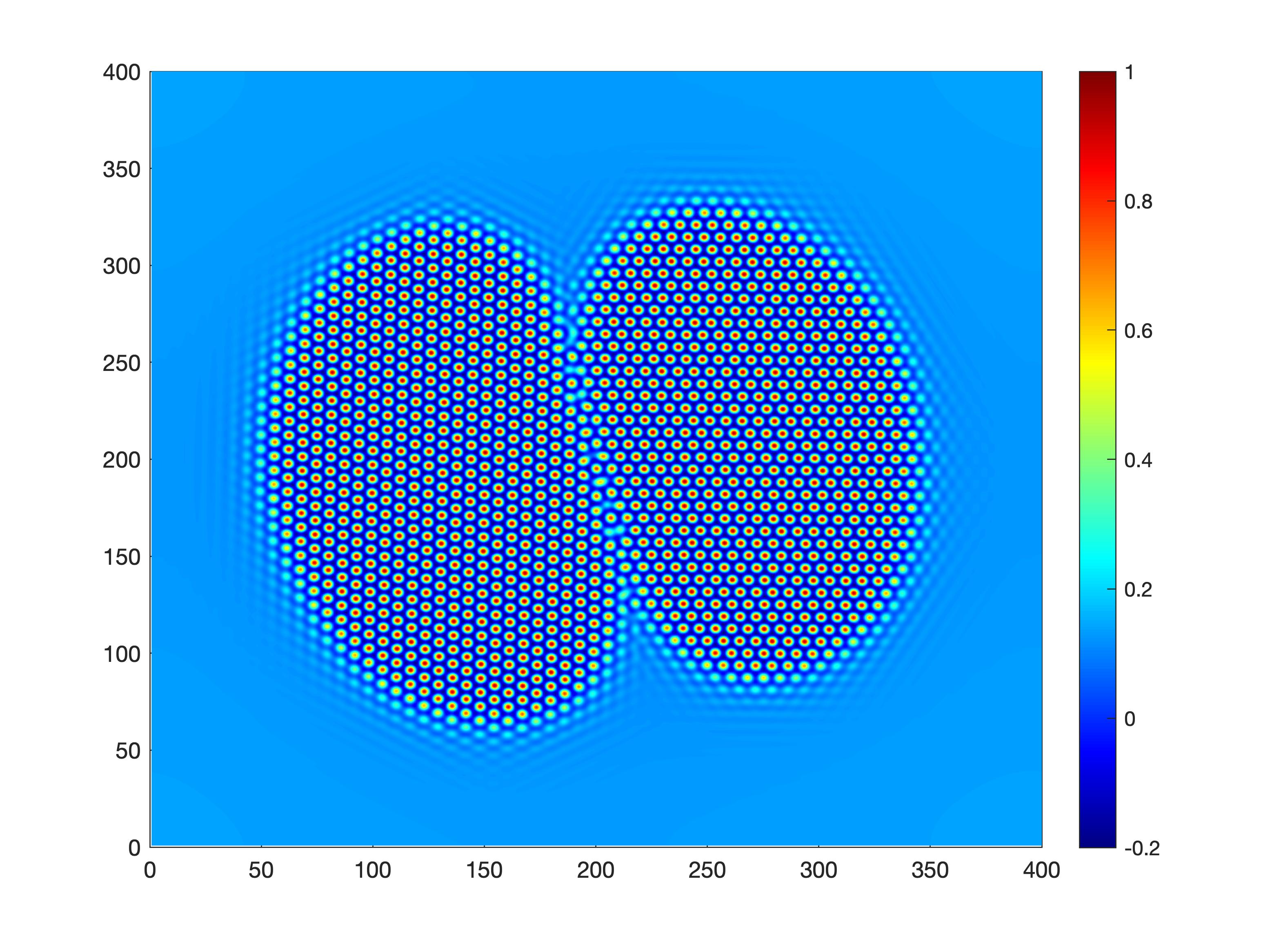}
\includegraphics[width = 2.6in]{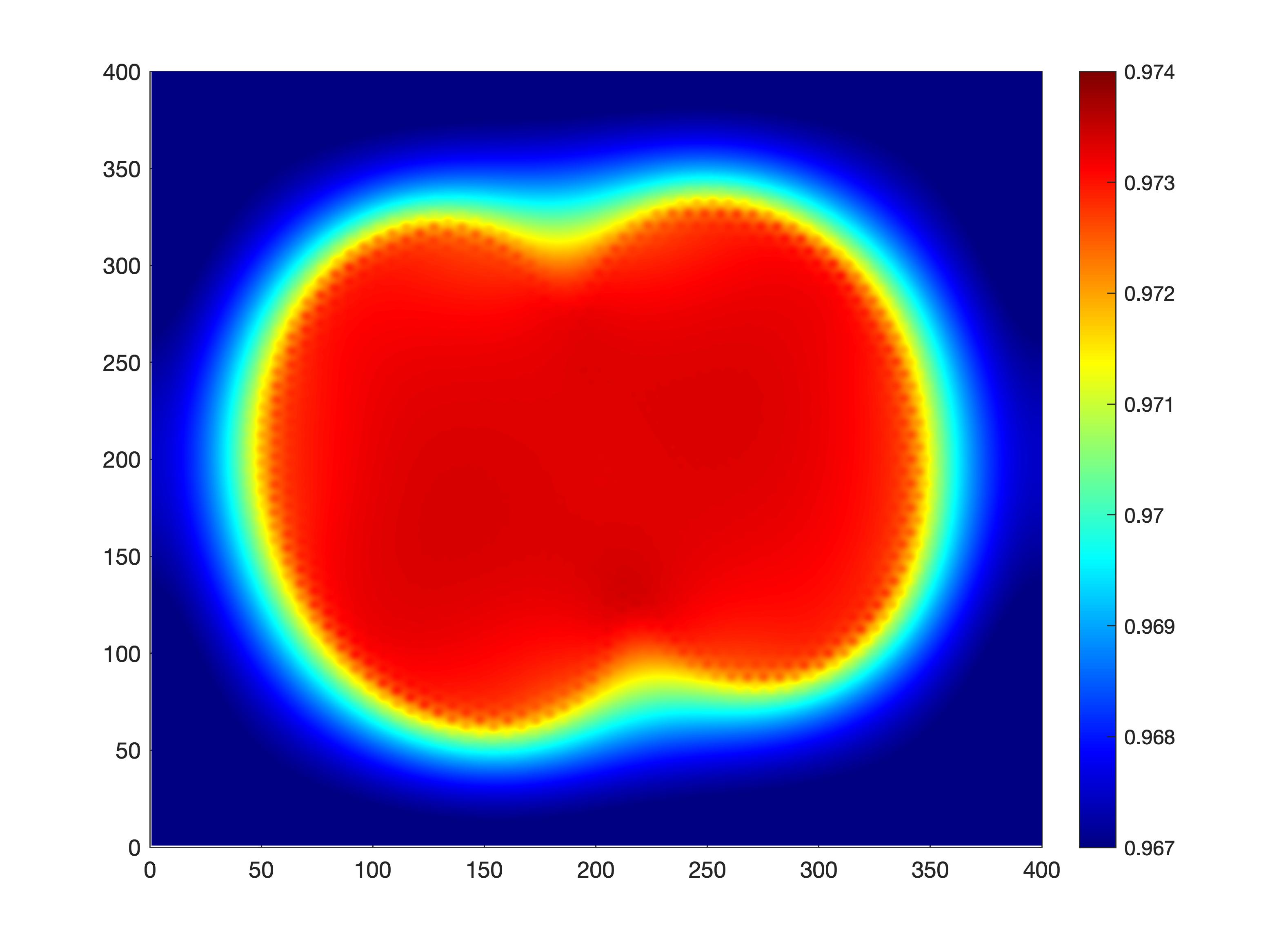}
	\end{center}
\caption{A unary crystal growing in its supersaturated liquid phase: (left column) the $\psi$ field and (right column) the temperature field. Parameters for the test are given in the caption of Figure~\ref{fig:melting-exact-2}. In addition, $\Omega = (0,400)^2$ and $M = 0.1$.}
	\label{fig:solid-sim}
	\end{figure}

For the numerical solution, we use a Fourier pseudo-spectral discretization of space, coupled with a stabilized, linear, first-order IMEX algorithm for time discretization. This scheme is not designed to keep the temperature field positive, and will not guarantee entropy production. More sophisticated schemes will be developed in the future that guarantee these properties theoretically. See, for example, the related numerical works of the PFC model with constant temperature~\cite{dong18a, hu09, wise09a}, the modified PFC equation~\cite{baskaran13a, baskaran13b}, the square PFC equation~\cite{Cheng2019d}. We perform a singe test, the results for which are shown in Figure~\ref{fig:solid-sim}. The parameters for the test are given in the captions of Figure~\ref{fig:melting-exact-2}, where we give a cartoon description of the free energy landscape for the parameters.

We seed the center of the domain with a two-grain crystal that is very near to its equilibrium state at $T=1.0$, with an average density of $A = 0.16$. (See Figure~\ref{fig:melting-exact-2}(left).) The liquid phase surrounding the crystal seed is supersaturated, meaning its average density (at $A = 0.14$) is higher than that of the equilibrium state (just below $A = 0.11$, see, again, Figure~\ref{fig:melting-exact-2}(left)). The initial temperature is roughly $T = 0.97$. This shifts the equilibria slightly; see Figure~\ref{fig:melting-exact-2}(right). The new equilibrium average would be computed using the Maxwell common-tangent construction, of course. The extra mass is ejected from the liquid and attaches to the solid seed, and the seed grows. As it does the temperature changes due the the release of latent heat, and we see a commonly observed Gibbs-Thompson-like temperature jump effect at the boundary of the crystal.

	\section*{Acknowledgements}
	
This work is partially supported by the National Science Foundation (USA) grants  NSF DMS-2012669 (C.~Wang) and NSF DMS-1719854, DMS-2012634 (S.M.~Wise). SMW thanks Ken Elder for bringing reference~\cite{Kocher2019} to his attention and for several discussions about PFC models.
		
	\bibliographystyle{plain}
	\bibliography{PFC}
	\end{document}